\documentclass[paper,11pt]{article}

\makeatletter
\@addtoreset{equation}{section}
\makeatother

\usepackage{etoolbox}
\makeatletter
\patchcmd\end@float{\@cons\@currlist\@currbox}
   {\@cons\@currlist\@currbox
    \global\holdinginserts\@ne}
    {}{}
    
\apptocmd\@specialoutput{\global\holdinginserts\z@}

\makeatother

\usepackage{cite}
\usepackage{epsfig,amsfonts}
\usepackage{graphicx}
\usepackage{caption}
\usepackage{subcaption}
\usepackage{amsmath}
\usepackage{amsthm,amssymb}
\usepackage{mathrsfs}
\usepackage{hhline}
\usepackage{braket}
\usepackage{upgreek}
\usepackage[ruled,vlined]{algorithm2e}
\usepackage[linktocpage=true]{hyperref}
\usepackage[dvipsnames]{xcolor}
\usepackage{musicography}
\usepackage[normalem]{ulem}
\definecolor{purple_nice}{rgb}{0.4,0.2,0.7}
\definecolor{fuel_blue}{RGB}{42,162,185}
\definecolor{YInMn_blue}{RGB}{46, 80, 144}
\definecolor{ultramarine}{RGB}{63, 0, 255}
\definecolor{KLEIN_blue}{rgb}{0, 0.18, 0.65}
\hypersetup{
    colorlinks=true,
    linkcolor=YInMn_blue,
    citecolor=ultramarine,
    filecolor=fuel_blue,
    urlcolor=KLEIN_blue,
}

\renewcommand{\Re}{\operatorname{Re}}
\renewcommand{\Im}{\operatorname{Im}}
\usepackage{enumitem}

\def\be{\begin{equation}}
\def\ee{\end{equation}}
\def\bea{\begin{eqnarray}}
\def\eea{\end{eqnarray}}

\def\XXint#1#2#3{{\setbox0=\hbox{$#1{#2#3}{\int}$}
    \vcenter{\hbox{$#2#3$}}\kern-.5\wd0}}

\newcommand{\TTB}{\mathrm{T}\overline{\mathrm{T}}}
\newcommand{\tqt}{\tilde{q}_t}
\newcommand{\beq}{\begin{equation}}
\newcommand{\eeq}{\end{equation}}

\def\({\left(}
\def\){\right)}
\def\[{\left[}
\def\]{\right]}
\newcommand{\e}{\epsilon}
\begin{document}
\begin{titlepage}

\title{{\huge  The Generalized Born Oscillator
and the Berry–Keating Hamiltonian}}
\author{Francesco Giordano$^{1\musFlat}$, Stefano Negro$^{2\musNatural}$ and Roberto Tateo$^{1\musSharp}$\\[0.3cm]}
\date{\footnotesize{$^1$ Dipartimento di Fisica,\\ Università di Torino and INFN, Sezione di Torino, Via P. Giuria 1, 10125, Torino, Italy \\[0.1cm]$^2$ Center for Cosmology and Particle Physics, New York University, New York,\\ NY 10003, U.S.A.\\[0.3cm] $^{\musFlat}$\texttt{\href{mailto:giordano.f413@gmail.com}{giordano.f413@gmail.com},} $^{\musNatural}$\texttt{\href{mailto:stefano.negro@nyu.edu}{stefano.negro@nyu.edu},} $^{\musSharp}$\texttt{\href{mailto:roberto.tateo@unito.it}{roberto.tateo@unito.it}}\\}}
\maketitle

\begin{abstract}
In this study, we introduce and investigate a family of quantum mechanical models in 0+1 dimensions, known as generalized Born quantum oscillators. These models represent a one-parameter deformation of a specific system obtained by reducing the Nambu-Goto theory to 0+1 dimensions. Despite these systems showing significant similarities with $\TTB$-type perturbations of two-dimensional relativistic models, our analysis reveals their potential as interesting regularizations of the Berry-Keating theory.
We quantize these models using the Weyl quantization scheme up to very high orders in $\hbar$. By examining a specific scaling limit, we observe an intriguing connection between the generalized Born quantum oscillators and the Riemann-Siegel $\theta$ function.

\end{abstract}
\end{titlepage}
\newpage
\tableofcontents
\newpage
\section{Introduction}
\label{sec:intro}

In the early 20th century, Hilbert and Pólya speculated about the imaginary parts of the complex zeros of the Riemann zeta function. They proposed that a self-adjoint operator $H$ could have these imaginary parts as eigenvalues; discovering such an operator would confirm the famous Riemann hypothesis.
Assuming the Riemann hypothesis is valid, Montgomery and Odlyzko deduced that the local statistical behavior of the Riemann roots resembles the Gaussian unitary ensemble (GUE) in random matrix theory. Building on this insight, Berry \cite{BerryChaos} suggested the likely existence of a classical Hamiltonian system with chaotic behavior and isolated periodic prime-number-related orbits. The quantum theory's spectrum from this system would reveal the complex Riemann roots, with their GUE statistics implying time-reversal symmetry violation by the Hamiltonian.

In 1999, Connes \cite{Connes1998TraceFI}, Berry and Keating \cite{Berry1999TheRZ} explored a semiclassical model involving a one-dimensional particle with a classical Hamiltonian $H_{\mathrm{BK}} = p q$. This Hamiltonian breaks time-reversal symmetry, $H_{\mathrm{BK}}(p,q)= - H_{\mathrm{BK}}(-p,q)$ and the classical orbits in this model are unbounded hyperbolas in phase space. By imposing boundary conditions, the Hamiltonian $H_{\mathrm{BK}}$ is ``regularized'' to a well-defined Hermitian operator with a discrete spectrum. In this setting, the count of  states with positive energy less than $E = \hbar T$ is related to the leading asymptotic contributions to the function $\overline{N(T)}$, measuring the average number of complex zeroes of the Riemann zeta function with positive imaginary part less than $T$. The approaches of Berry and Keating and of Connes differ in the way the classical Hamiltonian is regularized and, consequently, on how the asymptotics of $\overline{N(T)}$ is recovered: respectively as the number of states present in the spectrum \cite{Berry1999TheRZ} and as those missing from the continuum \cite{Connes1998TraceFI}.

In 2008, Sierra and Townsend \cite{Sierra:2008se} put forward a physical realization of the Hamiltonian $H_{\mathrm{BK}}$, by considering the lowest Landau level limit of a quantum-mechanical model. The model describes a charged particle moving on a planar surface subjected to a static electrostatic potential $V= V_0 \,xy$ and a uniform magnetic field $B$, perpendicular to the $x-y$ plane.
Various variants of the Berry-Keating Hamiltonian have been introduced and studied \cite{Sierra2011, Sierra_2019}, including one that incorporates broken time-reversal symmetry and closed orbits \cite{Berry2011}. Studying these models and their quantization has its own physical and mathematical significance, not necessarily tied to the initial proposal associated with the Hilbert and Polya's conjecture.

This work introduces a novel two-parameter  \emph{time-reversal symmetric}  regularization of the Berry-Keating model -- the \emph{Generalized Born oscillator} -- which incorporates naturally the regularization prescription of Connes. The Generalized Born oscillator is a system with closed trajectories, and its quantization does not necessitate a regularization prescription. Thus, by using Weyl's quantization technique, we are able to extract the number of states $N(E)$ beyond the leading semiclassical approximation to an in-principle arbitrary order in $\hbar$. As an interesting by-product,  we will see that this expansion reproduces that of $\overline{N(T)}$, together with correction terms that vanish as one of the system's parameters is sent to $\infty$. Additionally, the asymptotics of the number of states displays a term linear in $T$ which accounts for the fact that our model can be interpreted as a regularization of Connes boundary conditions.

This work is structured as follows.  \S\ref{sec:BK} contains a short review of the known results on the Berry-Keating theory. The \emph{Born oscillator}, a special case of the Generalized Born oscillator, will be introduced in \S\ref{sec:BO} as a dimensional reduction of relativistic  free scalar theory in $1+1$ dimensions, deformed with the $\mathrm{T}\overline{\mathrm{T}}$ operator. In this same section, we will describe the properties of the Born oscillator and study the spectrum arising from its Weyl quantization, comparing the state-counting function $N(E)$ to the function $\overline{N(T)}$. 
We will see that for large $E = \hbar T$ the latter is reproduced order by order, although it is accompanied by a number of spurious terms that cannot be eliminated naturally. Furthermore, in \S\ref{sec:GBO} we introduce an additional parameter $u$ in the system, defining the Generalized Born oscillator. Studying its properties and its spectrum, we find that this system constitutes a regularization of Connes' regularization for the Berry-Keating theory, and it is such that in the $u\rightarrow\infty$ limit, the large $E$ behaviors of $N(E)$ and $\overline{N(T)}$ coincide, up to a term linear in $E$. In \S\ref{sec:conclusions} we conclude and give some outlook. In addition, there  are five Appendices. In Appendix \ref{sec:RZF}, we give a concise overview of counting functions and the Riemann-Siegel $\theta$-function, with a focus on their emergence in the context of the Riemann $\zeta$ function, which initially inspired the work of Berry and Keating. Some technical details about the Weyl quantization are collected in Appendix \ref{app:Weyl_quant}, while Appendix \ref{app:iterative} describes the procedure that we employed to determine the spectrum. Appendix \ref{app:Weyl_WKB} contains a comparison of the Weyl quantization with the more familiar WKB approach. Finally, in Appendix \ref{app:poly} we collect some lengthy expressions pertaining to the $\hbar\rightarrow 0$ expansions of the quantization conditions for the Born oscillator and for the Generalized one.

\section{The Berry–Keating Theory}
\label{sec:BK}
\begin{figure}[t!]
\centering
\includegraphics[width=0.6\textwidth]{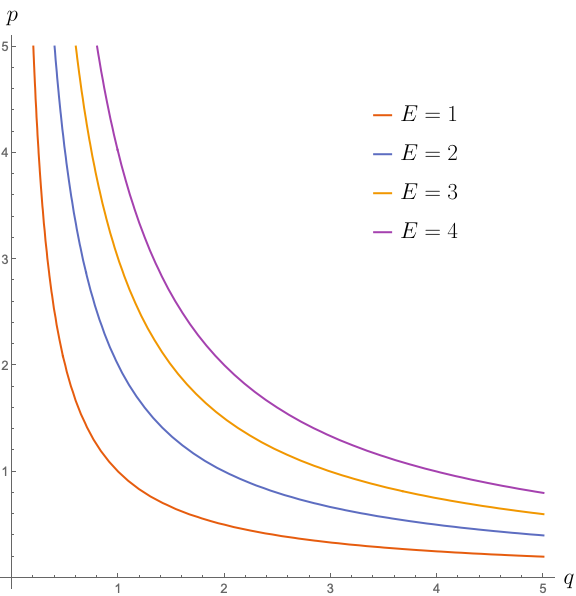}
\caption{\label{fig:pq} Classical trajectories $E=pq$ for $E=1,2,3,4$.}
\end{figure}
In \cite{Berry1999TheRZ}, M. Berry and J. Keating studied the strikingly simple classical Hamiltonian
\beq
    H_{\mathrm{BK}}(p,q) = p q\;,
\label{eq:BK_hamiltonian}
\eeq
hereafter referred to as the \emph{Berry-Keating Hamiltonian}, in relation to the smooth part $\overline{N(T)}$ of the counting function \eqref{eq:average_numb_zeroes}. 
In particular, they proposed that the counting function for the spectrum of an appropriate quantization of \eqref{eq:BK_hamiltonian} will reproduce $\overline{N(T)}$ in the large $T$ limit.

The Hamiltonian \eqref{eq:BK_hamiltonian} generates a flow on the $2$-dimensional phase space $(p,q)$ whose classical trajectories are branches of hyperbolas (see fig. \ref{fig:pq})
\beq
    H_{\mathrm{BK}}(p,q) = p q = E\;.
\label{eq:BK_trajectories}
\eeq
Since the trajectories are open, the associated quantum mechanical system will have a continuum spectrum. This, of course, is problematic if one's goal is to establish a  link between the distribution of the spectrum of (\ref{eq:BK_trajectories}) and $\overline{N(T)}$. In order to overcome this problem, we can proceed in two complementary ways. The first, proposed by Berry and Keating \cite{Berry1999TheRZ}, is to impose the following constraints
\beq
    \vert q \vert \geq l_q\;,\qquad \vert p \vert \geq l_p\;,\qquad l_{(p,q)} > 0\;.
\label{eq:BK_constraints}
\eeq
\begin{figure}[h!]
\centering
\includegraphics[width= 0.6\textwidth]{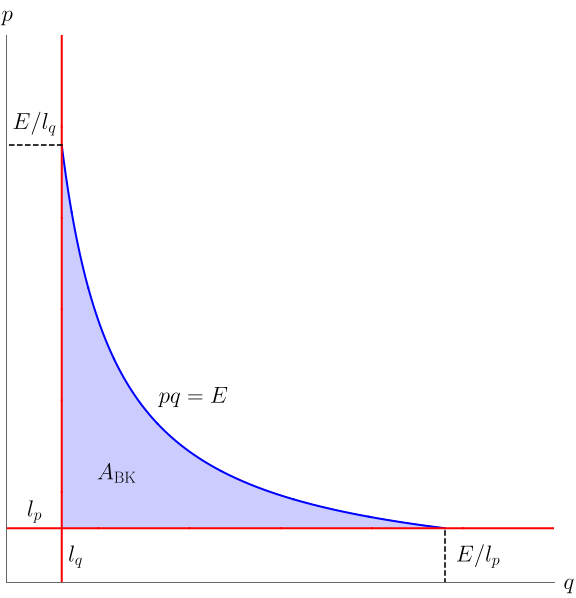}
\caption{\label{fig:pq_l}Area of the phase space with the constraints $|q| \geq l_q$, $|p| \geq l_p$. $A_{BK}$ is the area enclosed by the curves $E=pq$, $q=l_q$ and $p=l_p$. }
\end{figure}
With these cutoffs, the trajectories are now bounded (see fig. \ref{fig:pq_l}) and it is possible to employ semi-classical analysis to evaluate the number of states $N_{\mathrm{BK}}$ as the area $A_{\mathrm{BK}}$ enclosed by the curve $p q = E$ and the lines $p=l_p$, $q=l_q$, in units of Planck cells ($2\pi \hbar$) and corrected by the \emph{Maslov index}\footnote{The Maslov index is what gives the $1/2$ correction in the quantization of the harmonic oscillator.} $i_{\mathrm{M}} = -1/8$ (see \cite{8e50c801912c4798a2696cbe087fdcf6} for details)
\beq
    N_{\mathrm{BK}} = \frac{A_{\mathrm{BK}}}{2\pi\hbar} + i_{\mathrm{M}} = \frac{E}{2\pi\hbar}\left(\log\frac{E}{l_q l_p} - 1\right) + \frac{l_q l_p}{2\pi\hbar} - \frac{1}{8}\;.
\label{eq:BK_num_states}
\eeq
It is a straightforward check that with the choice
\beq
    E = \hbar T\;,\qquad l_p l_q = 2\pi \hbar\;,
\label{eq:BK_parameter_identification}
\eeq
the first two terms of the expansion \eqref{eq:sieg_asy} are exactly reproduced.

The other possibility to regularize the semi-classical spectrum of \eqref{eq:BK_hamiltonian} was proposed by A. Connes \cite{Connes1998TraceFI}. In this case, one imposes a large cut-off $\Lambda$ on both $q$ and $p$
\begin{figure}[h!]
\centering
\includegraphics[width= 0.6\textwidth]{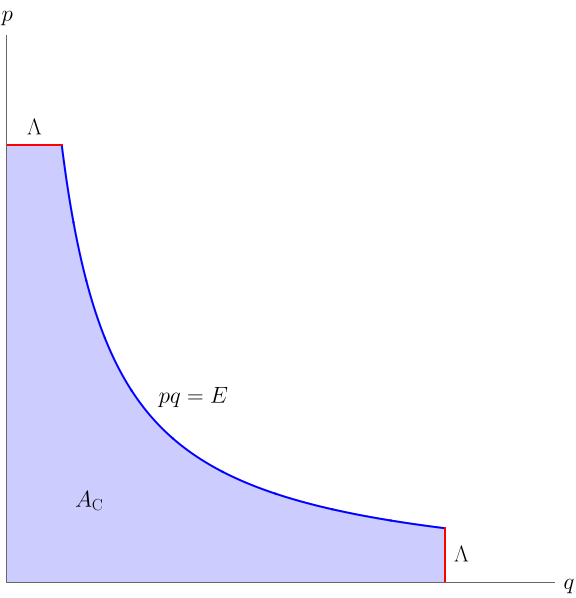}
\caption{\label{fig:pq_lambda} Phase space with the constraints $|q| \leq \Lambda $, $|p| \leq \Lambda$. The area is $A_{C}$.}
\end{figure}
\beq
    \vert q \vert \leq \Lambda\;,\qquad \vert p \vert \leq \Lambda\;,\qquad \Lambda > 0\;.
\label{eq:C_constraints}
\eeq
Here too, the trajectories become bounded (see fig. \ref{fig:pq_lambda}) and the number of states $N_{\mathrm{C}}$ in semi-classical quantization reads
\beq
    N_{\mathrm{C}} = \frac{A_{\mathrm{C}}}{2\pi\hbar} = \frac{E}{2\pi\hbar}\log\frac{\Lambda^2}{2\pi\hbar} - \frac{E}{2\pi\hbar}\left(\log\frac{E}{2\pi\hbar} - 1\right)\;.
\label{eq:C_spectrum_lambda}
\eeq
This expression contains two terms, one which diverges as $\Lambda\rightarrow\infty$, corresponding to the continuum spectrum, and a second one reproducing to the leading asymptotic behavior of $\overline{N(T)}$ with $E=\hbar T$ and a negative sign overall. Due to their expressions, $N_{\mathrm{C}}$ and $N_{\mathrm{BK}}$ are often referred to as the \emph{absorption} and \emph{emission spectrum}, respectively. Let us notice that, by choosing the cutoff $\Lambda$ to be
\beq
    \Lambda = \frac{E}{\sqrt{l_p l_q}} = \frac{E}{\sqrt{2\pi\hbar}}\;,
\label{eq:Connes_cutoff}
\eeq
we can rewrite
\beq
    N_{\mathrm{C}} = \frac{E}{2\pi\hbar} \left(\log \frac{E}{2\pi\hbar} + 1\right) = \frac{E}{2\pi\hbar} \left(\log \frac{E}{2\pi\hbar} - 1\right) + 1 + \frac{2E-2\pi\hbar}{2\pi\hbar}\;.
\label{eq:C_spectrum_no_lambda}
\eeq
We recognize in the last addend the area $2E - l_p l_q$ (with $l_p l_q = 2\pi\hbar$) of the two rectangles that were omitted in the Berry-Keating regularization \eqref{eq:BK_constraints} (see fig. \ref{fig:pq_l}). The addition of the Maslov index $i_{\mathrm{M}} = -1/8$ reproduces the $7/8$ constant term of \eqref{eq:sieg_asy}.

The derivation of the spectra $N_{\mathrm{BK}}$ and $N_{\mathrm{C}}$ are quite simple and far from being formal. However, Berry and Keating in their work \cite{8e50c801912c4798a2696cbe087fdcf6}, remarked that the fact they reproduce the leading asymptotic of $\overline{N(T)}$ is unlikely to be a coincidence. They also emphasized the necessity of replacing the semiclassical regularization \eqref{eq:BK_constraints}—or \eqref{eq:C_constraints} for that matter—with a procedure that naturally generates a discrete spectrum.

One of the main results of this  paper is the introduction of such a procedure, in the form of a family of deformations of the Berry-Keating Hamiltonian \eqref{eq:BK_hamiltonian}. As we will see, the parameters controlling these deformations serve as natural regulators of the trajectories \eqref{eq:BK_trajectories} and the quantum spectrum, obtained with the Weyl quantization of the classical Hamiltonian, is such that the subleading terms of the asymptotic expansion \eqref{eq:sieg_asy} are reproduced in a limiting regime. 

\section{The Born Oscillator}
\label{sec:BO}
In order to formulate a deformation of the Berry-Keating Hamiltonian \eqref{eq:BK_hamiltonian} we start, perhaps counterintuitively, from $1+1$ space-time dimensions. Let us consider the Nambu-Goto model in the static gauge with a single transversal field $\varphi(t,x)$, describing the fluctuations of a $2$-dimensional surface embedded in $3$-dimensional space-time. The Lagrangian of this model is
\beq
    L_{\mathrm{NG}} = \int dx\, \mathscr{L}_{\mathrm{NG}} = \int dx\, \frac{1 - \sqrt{1+\lambda \varphi'(t,x)^2 - \lambda \dot{\varphi}(t,x)^2}}{\lambda}\;,
\label{eq:NG_lagrangian}
\eeq
where $\mathscr{L}_{\mathrm{NG}}$ is the Lagrangian density and the primes $\varphi'(t,x)$ and dots $\dot{\varphi}(t,x)$ denote derivations with respect to $x$ and $t$, respectively. As shown in \cite{Cavaglia:2016oda}, the model described by \eqref{eq:NG_lagrangian} can be considered as a $\mathrm{T}\overline{\mathrm{T}}$ deformation of a free massless scalar field in $1+1$ dimensions. From this point of view, $\lambda$ -- that in the string theory perspective is the inverse of the string tension -- plays the role of a deformation parameter and the theory is defined by a flow equation for the action
\beq
    \begin{split}
        &\frac{d}{d\lambda} L_{\mathrm{NG}} = -\frac{1}{2} \int dx\, \det_{\mu\nu} T_{\mu\nu}(t,x)\;,\\ &\lim_{\lambda\rightarrow 0} L_{\mathrm{NG}} = \int dx\, \frac{\dot{\varphi}(t,x)^2 - \varphi'(t,x)^2}{2}\;,
    \end{split}
\label{eq:TTbar_deformation}
\eeq
where $T_{\mu\nu}(t,x)$ is the canonical energy-momentum tensor of the theory described by $L_{\mathrm{NG}}$. Performing a Legendre transformation, we readily derive the Hamiltonian of the model
\beq
    H_{\mathrm{NG}} = \int dx\,\frac{\sqrt{1 + \lambda \pi(x)^2}\sqrt{1 + \lambda \varphi'(x)^2} - 1}{\lambda}\;,
\label{eq:NG_hamiltonian}
\eeq
which obeys the following limiting behaviors
\beq
    \lim_{\lambda\rightarrow 0} H_{\mathrm{NG}} = \int dx\, \frac{\pi(x)^2 + \varphi'(x)^2}{2}\;,\qquad \lim_{\lambda\rightarrow \infty} H_{\mathrm{NG}} = \int dx\, \vert\pi(x)\vert\,\vert \varphi'(x)\vert\;.
\label{eq:NG_limits}
\eeq
We see that the  $\mathrm{T} \overline{\mathrm{T}}$ deformation \eqref{eq:TTbar_deformation} determines a one-parameter family of models that interpolate -- at least at the classical level -- from a theory of a free massless scalar to one with vanishing Lagrangian. The latter is suggestively similar to the Berry-Keating theory \eqref{eq:BK_hamiltonian}, albeit in $1+1$ dimensions instead of $1+0$.

One way to move from a field theory to quantum mechanics is to perform a standard dimensional reduction, compactifying the space dimension on a circle of vanishing radius. This route was followed in \cite{Gross:2019ach}, where a whole family of ``$\mathrm{T}\overline{\mathrm{T}}$-like'' flows was introduced:
\beq
    \frac{d}{d\lambda} H_{\lambda} = f(H_{\lambda})\;.
\label{eq:Gross_TTbar_QM}
\eeq
These flows are such that the deformed Hamiltonian $H_{\lambda}$ is a function of the undeformed one $H_{\lambda} = \mathcal{H}_{\lambda}(H_0)$. As a consequence, the theories determined by \eqref{eq:Gross_TTbar_QM} are not of interest for our purposes, since they cannot yield deformations of the Berry-Keating theory \eqref{eq:BK_hamiltonian} with compact trajectories\footnote{In fact if we want \eqref{eq:BK_hamiltonian} to be reproduced by $H_{\lambda} = \mathcal{H}_\lambda(H_0)$ at some point of the flow $H_{\lambda_k} = H_{\mathrm{BK}}$, we must have $H_{\lambda} = \mathcal{F}_{\lambda}(H_{\mathrm{BK}})$, where $\mathcal{F}_{\lambda} = \mathcal{H}_{\lambda}\circ \mathcal{H}_{\lambda_k}^{-1}$. Then the trajectories will have the form $p q = \mathcal{F}_{\lambda}^{-1}(E)$, which are clearly open.}. We instead follow a dual route and discretize the space direction on a lattice with spacing $\Delta x$. Using the notation $f_n = f(n \Delta x)$ for any function $f$ and choosing a symmetric version of the first derivative squared, we can write the Hamiltonian
\eqref{eq:NG_hamiltonian} as
\beq
    H_{\mathrm{NG}} = \sum_{n\in\mathbb{Z}} \Delta x\,\frac{\sqrt{1 + \lambda\pi_n^2} \sqrt{1 + \lambda\frac{\left(\varphi_{n+1} - \varphi_n\right)^2}{2\Delta x^2} + \lambda\frac{\left(\varphi_n - \varphi_{n-1}\right)^2}{2\Delta x^2}} - 1}{\lambda}\;.
\eeq
This can be interpreted as describing a system of infinitely many particular oscillators, coupled by some potential:
\beq
    H_{\mathrm{NG}} = \sum_{n\in\mathbb{Z}} \Delta x \left[H_{n} + V_n\right]\;,
\eeq
where
\beq
    H_{n} = \frac{\sqrt{1 + \lambda \pi_n^2} \sqrt{1 + \lambda \frac{\varphi_n^2}{\Delta x^2}} - 1}{\lambda}\;,
\label{eq:BO_multiple}
\eeq
and
\beq
    V_n = \sqrt{1 + \lambda \pi_n^2} \sum_{m=1}^\infty \left(\begin{array}{c} 1/2 \\ m \end{array}\right) \frac{\lambda^{m-1}}{2^m \Delta x^{2m}} \frac{\left(\varphi_{n+1}^2 + \varphi_{n-1}^2 - 2 \varphi_n\left(\varphi_{n+1} + \varphi_{n-1}\right)\right)^m}{\left(1 + \lambda \varphi_n^2\right)^{\frac{2m-1}{2}}}\;.
\eeq
Isolating a single site, say $n=0$, and setting $\pi_0 = p$, $\varphi_0 = \Delta x q$, we obtain a theory determined by the Hamiltonian
\beq
    H_{\mathrm{BO}} = \frac{\sqrt{\left(1 + \lambda p^2\right)\left(1 + \lambda q^2\right)}}{\lambda}\;,
\label{eq:BO_Hamiltonian}
\eeq
where we discarded the ``cosmological constant'' term $-1/\lambda$. Following \cite{CoppaThesis}, \footnote{See \cite{CoppaArticle}, for some complementary results on this interesting quantum mechanical model.} we will call the model determined by $H_{\mathrm{BO}}$ the \emph{Born oscillator}. It is interesting to remark that, while \eqref{eq:BO_Hamiltonian} clearly does not satisfy a flow equation of the form \eqref{eq:Gross_TTbar_QM} proposed in \cite{Gross:2019ach}, the corresponding Lagrangian
\beq
    L_{\mathrm{BO}} = -\frac{\sqrt{1 + \lambda q^2 - \lambda \dot{q}^2}}{\lambda}\;,
\eeq
obeys the flow equation
\beq
    \frac{d}{d\lambda} L_{\mathrm{BO}} = \frac{1 + \lambda^2 L_{\mathrm{BO}}^2}{2 \lambda L_{\mathrm{BO}}}\;.
\eeq
We plan to study in detail theories determined by flow equations of this kind in a future work.

\begin{figure}[t!]
\centering
\includegraphics[width= 0.6\textwidth]{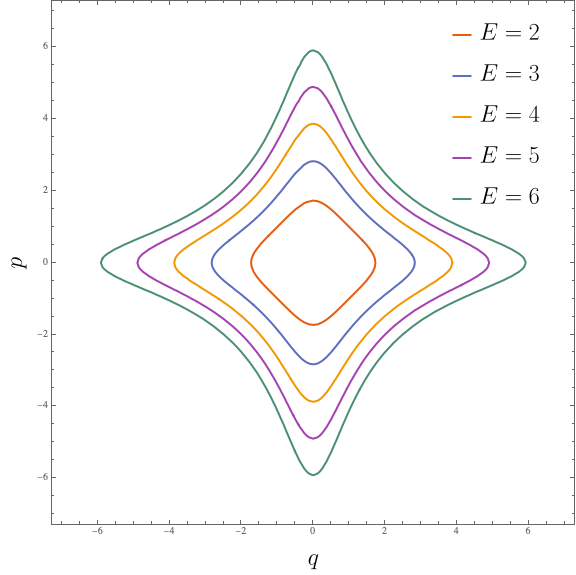}
\caption{\label{fig:Born} Some classical trajectories of the Born oscillator $H_{\mathrm{BO}}=E$ for fixed $\lambda = 1$.}
\end{figure}

\begin{figure}[t!]
\centering
\includegraphics[width= 0.6\textwidth]{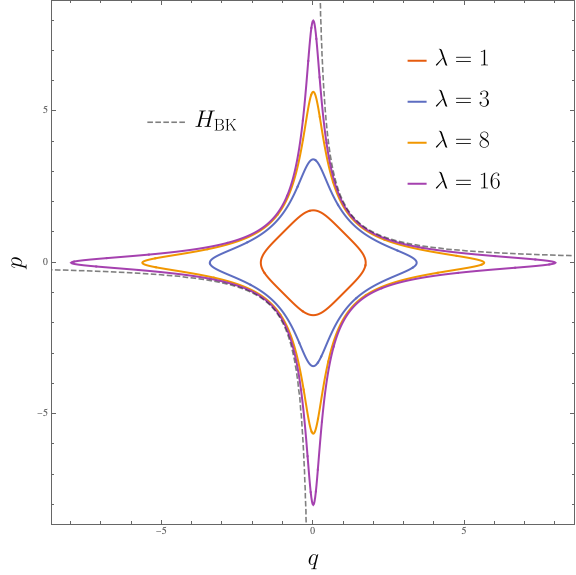}
\caption{\label{fig:Born_vs_BK} Some classical trajectories of the Born oscillator $H_{\mathrm{BO}}=E$ for fixed $E = 2$.}
\end{figure}

Just as it happened in Nambu-Goto \eqref{eq:NG_limits}, by dialing the deformation parameter $\lambda$, the Born oscillator \eqref{eq:BO_Hamiltonian} interpolates between a standard harmonic oscillator (up to a $1/\lambda$ term) and a theory with vanishing Lagrangian which is nothing else but the Berry-Keating one \eqref{eq:BK_hamiltonian}:
\beq
    \lim_{\lambda\rightarrow 0} \left(H_{\mathrm{BO}} - \frac{1}{\lambda}\right) = \frac{p^2 + q^2}{2} = H_{\mathrm{HO}}\;,\qquad \lim_{\lambda\rightarrow\infty} H_{\mathrm{BO}} = \vert p q \vert = \vert H_{\mathrm{BK}} \vert\;.
\eeq
Importantly, by studying the classical trajectories of the Born oscillator \eqref{eq:BO_Hamiltonian}, we see that they are \emph{closed} for any finite value of $\lambda$  and approach the Berry-Keating ones as this parameter grows larger (see figures \ref{fig:Born} and \ref{fig:Born_vs_BK}).
Another important fact that we notice is that the hyperbolic classical trajectories \eqref{eq:BK_trajectories} of the Berry-Keating theory are also approached as the energy grows, in accordance with the limiting behaviors
\beq
    \begin{split}
        H_{\mathrm{BO}} &\underset{{\vert p\vert ,\vert q\vert\rightarrow \infty}\atop{p q>0}}{\sim} H_{\mathrm{BK}} + \frac{p q}{2\lambda} \left(\frac{1}{p^2} + \frac{1}{q^2}\right) + \cdots\;, \\
        H_{\mathrm{BO}} &\underset{{\vert p\vert ,\vert q\vert\rightarrow \infty}\atop{p q<0}}{\sim} -H_{\mathrm{BK}} - \frac{p q}{2\lambda} \left(\frac{1}{p^2} + \frac{1}{q^2}\right) + \cdots\;.
    \end{split}
\eeq
Finally, from both the expression \eqref{eq:BO_Hamiltonian} and the figures \ref{fig:Born} and \ref{fig:Born_vs_BK} it is clear that our systems possess a $D_4$ dihedral symmetry. Thus, we can consider $H_{\mathrm{BO}}$ as a ``$D_4$ regularization'' of the Berry-Keating model.

The properties of the Hamiltonian \eqref{eq:BO_Hamiltonian} we just described raise the hope that its quantum spectrum may indeed reproduce the expansion \eqref{eq:sieg_asy} at large energies. We can get a first confirmation of this by evaluating the number of states $N_{\mathrm{BO}}$ at the first semi-classical order:
\beq
    N_{\mathrm{BO}} + \frac{1}{2} = \frac{1}{2\pi \hbar} \iint\limits_{H_{\mathrm{BO}} \leq E} dp dq + \mathcal{O}(\hbar)\;.
\eeq
With some simple manipulations, we arrive at the expression
\beq
    N_{\mathrm{BO}} + \frac{1}{2} = 2\frac{\left(1+\lambda q_t^2\right)\mathsf{K}(-\lambda q_t^2) - \mathsf{E}(-\lambda q_t^2)}{\pi\hbar\lambda} + \mathcal{O}(\hbar)\;,
\eeq
where $q_t = \sqrt{(\lambda^2 E^2-1)/\lambda}$ is the positive turning point of the trajectory at energy $E$ and $\mathsf{K}(x)$, $\mathsf{E}(x)$ are the Legendre complete elliptic integrals (see \S 19 in \cite{NIST:DLMF}). For large values of the spectral parameter $E$ we find the behavior
\beq
    N_{\mathrm{BO}} + \frac{1}{2} = 2\frac{E}{\pi \hbar}\left( \log(4E\lambda) - 1 \right) + \mathcal{O}(\hbar)\;.
\eeq
If we now perform the following identifications
\beq
    \begin{split}
        &N_{\mathrm{BO}} = 4 N(T) - 4\;, \\
        &E = \hbar T\;, \\
        &\lambda = \frac{1}{4}\frac{1}{2\pi\hbar}\;,
    \end{split}
\label{eq:parameter_identifications_1}
\eeq
we obtain the first two terms in \eqref{eq:sieg_asy}
\beq
    N(T) = \frac{T}{2\pi}\left(\log\left(\frac{T}{2\pi}\right) - 1\right) + \frac{7}{8} + \mathcal{O}\left(\frac{1}{T}\right)\;.
\label{eq:BO_semiclass_sieg}
\eeq
Let us provide some justification for the choices \eqref{eq:parameter_identifications_1}. The overall factor $4$ in front of $N(T)$ accounts for the $\mathbb{Z}_4$ symmetry of the Born oscillator trajectories. In fact, this symmetry divides the space of states of the theory into disjoint super-selection sectors, and we are counting only the states belonging to one of the sectors, the one connected to the fundamental state. Comparing with \eqref{eq:BK_parameter_identification}, we see that $\lambda$ is, up to a factor of $4$, the inverse of the cutoff $l_p l_q$ used in the Berry-Keating regularization \eqref{eq:BK_constraints}. Finally, the $-1$ shift of $N(T)$ is there because the lower energy state with energy $E=E_1= T_1/\hbar$ corresponds to $N_{\mathrm{BO}}(E_1)=0$, while we want $N(T)$ to be a counting function and thus satisfying $N(T_1)=1$.

While the expansion \eqref{eq:BO_semiclass_sieg} is an encouraging signal that the Born oscillator might indeed provide the sought-after trajectories of the Berry-Keating system, we need to devise a way to check that the sub-leading terms in the semi-classical expansion of the number of states $N_{\mathrm{BO}}$ agree with the expansion of the Riemann-Siegel $\theta$ function \eqref{eq:sieg_asy}. This task requires us to properly quantize the Hamiltonian $H_{\mathrm{BO}}$. The canonical procedure of promoting $(p,q)$ to operators, say in position space representation $(\hat{p},\hat{q}) = (-i\hbar\frac{d}{dq},q)$, is severely complicated by the fact that $H_{\mathrm{BO}}$ is non-polynomial in $p$ and $q$. One might think that expanding the square roots simplifies the problem, however, the resulting stationary Schr\"{o}dinger equation is a differential eigenvalue problem of infinite order, whose study is, to put it mildly, unwieldy. For this reason, we are going to employ a different approach, called Weyl quantization, to the evaluation of the number of states of the Born oscillator. This procedure is detailed in Appendix \ref{app:Weyl_quant}.

\subsection{Quantization of the Born Oscillator}
\label{subsec:QuantizationBO}
Using the Weyl quantization procedure to quantize the Born oscillator Hamiltonian \eqref{eq:BO_Hamiltonian} brings us to the following expansion (see Appendix \ref{app:iterative} for more details)
\beq
    n + \frac{1}{2} = \frac{1}{\hbar} \Sigma_0(E) + \sum_{m=1}^{\infty} \Sigma_m(E) \hbar^{2m-1}\;.
\label{eq:quant_cond_for_BO}
\eeq
The formulas needed to compute the coefficients $\Sigma_0$ and $\Sigma_m$ are given in \eqref{eq:Sigma_eq_final}. The first we already obtained in the previous section: it is $2/\pi$ times the area under the curve determined by the condition $H_{\mathrm{BO}}(p,q) = E$
\beq
    \Sigma_0(E) = 2\frac{\left(1 + \tqt^2\right)\mathsf{K}\left( -\tqt^2\right) - \mathsf{E}\left( -\tqt^2\right)}{\pi\lambda}\;,\qquad \tqt = \sqrt{\lambda^2 E^2 - 1}\;.
\label{eq:BO_sigma_0}
\eeq
The other coefficients require some more work. They all turn out to be of the following form
\beq
    \Sigma_m(E) = \lambda^{2m-1}\frac{P_m(\tqt^2)\mathsf{K}\left(-\tqt^2\right) + Q_m(\tqt^2)\mathsf{E}\left(-\tqt^2\right)}{\pi \tqt^{4m-2}\left(1+ \tqt^2\right)^{2m-1}}\;,
\label{eq:BO_sigma_general_form}
\eeq
where $P_m(x)$ and $Q_m(x)$ are polynomials of order $3m-2$. Here we display their expressions for $m=1,2$
\beq
    \begin{split}
        P_1(x) &= \frac{1 + x}{12}\;, \\
        Q_1(x) &= \frac{-1 + x}{12}\;,
    \end{split}
\label{eq:BO_sigma_elliptic_coefficients_1}
\eeq
\beq
    \begin{split}
        P_2(x) &= -\frac{56 + 233 x + 363 x^2 + 5323 x^3 + 2257 x^4}{2880}\;, \\
        Q_2(x) &= \frac{56 + 205 x + 264 x^2 + 5233 x^3 + 14 x^4}{2880}\;.
    \end{split}
\label{eq:BO_sigma_elliptic_coefficients_2}
\eeq
The polynomials for $m=3,4,5$ are reported in Appendix \ref{app:poly}, since they are quite cumbersome.

We are interested in the behavior of \eqref{eq:quant_cond_for_BO} for large $E$, equivalently large $\tqt$. For the elliptic integrals, we have
\beq
    \begin{split}
        \mathsf{K}(-\tqt^2) &= \frac{\log(4\tqt)}{\tqt} - \frac{\log(4\tqt) - 1}{4\tqt^3} + 3\frac{6\log(4\tqt) - 7}{128 \tqt^5} + \mathcal{O}\left(\frac{\log \tqt}{\tqt^7}\right)\;,\\
        \mathsf{E}(-\tqt^2) &= \tqt+\frac{2\log(4\tqt) + 1}{4\tqt} - \frac{4\log(4\tqt) - 3}{64\tqt^3} + 3\frac{\log(4\tqt) - 1}{128 \tqt^5} + \mathcal{O}\left(\frac{\log \tqt}{\tqt^7}\right)\;,
    \end{split}
\eeq
from which we can estimate the asymptotic behavior
\beq
    \Sigma_m(E) \underset{E\rightarrow\infty}{\sim} c\,E^{1-2m}\;.
\eeq
\subsection{The asymptotic behavior of the Born oscillator counting function}
Having explicit expressions for $m<6$, we can determine the large $E$ behavior of \eqref{eq:quant_cond_for_BO} up to order $\mathcal{O}\left(E^{-11}\right)$. Using the identifications \eqref{eq:parameter_identifications_1}, this takes the following form
\beq
    \begin{split}
        N(T) &= \frac{T}{2\pi}\left(\log\left(\frac{T}{2\pi}\right) - 1\right) + \frac{7}{8} + \frac{1}{48\pi T} + \frac{7}{5760\pi T^3} + \frac{31}{80640\pi T^5}\\
        &+\frac{127}{430080\pi T^7} + \frac{511}{1216512\pi T^9} - \mathrm{U.T.} + \mathcal{O}\left(\frac{\log T}{T^{11}}\right)\;,
    \end{split}
\label{eq:BO_large_E_numb_states}
\eeq
where we recognize the terms in the large $T$ expansion of the Riemann-Siegel $\theta$ function \eqref{eq:sieg_asy}. If the expansion \eqref{eq:BO_large_E_numb_states} looks too good to be true, that is because we hid all the ``unwanted terms'' inside $\mathrm{U.T.}$:
\beq
    \begin{split}
        \mathrm{U.T.} &= 8\pi \frac{\log\left(\frac{T}{2\pi}\right)}{T} + \pi\left(48\pi^2 - 1\right)\frac{6\log\left(\frac{T}{2\pi}\right) - 5}{3T^3}  \\
        &+ \pi \frac{300(5 - 32\pi^2 + 1034\pi^4)\log\left(\frac{T}{2\pi}\right) - (3477 - 12320 \pi^2 + 317440 \pi^4)}{120T^5} + \cdots\;.
    \end{split}
\label{eq:BO_spurious_terms}
\eeq
So we see that, while the Weyl quantization of the Born oscillator \eqref{eq:BO_Hamiltonian} does indeed seem to produce all the terms in the asymptotic expansion of the mean number of zeroes \eqref{eq:average_numb_zeroes}, it also generates several spurious terms. In principle, it is possible to eliminate at least part of them by performing a redefinition of $T$. For example, setting $T = T + \sum_{l=1}^{\infty} \alpha_{2l-1} T^{-2l+1}$ and carefully choosing the coefficients $\alpha_{2l-1}$ it is possible to eliminate all the terms $T^{-2l+1}\log T/(2\pi)$ in \eqref{eq:BO_spurious_terms}. However, this redefinition looks very unnatural and, what's more, fails to eliminate completely $\mathrm{U.T.}$ 

If the goal is to obtain a spectrum whose large energy expansion exactly reproduces the behavior of the Riemann-Siegel $\theta$ function, a more radical step should be taken: deform the Born oscillator by introducing an additional parameter $u$ and tune it appropriately to eliminate the ``spurious" terms. Before proceeding in this direction, it's worth noting that the results obtained in this section using the iterative procedure detailed in Appendix \ref{app:iterative} can also be reproduced by applying the more familiar WKB quantization method. This involves applying this standard procedure to an operator derived from the Hamiltonian $H_{\mathrm{BO}}$ by imposing a specific ordering that follows from the Weyl quantization prescription. For further details on this point, we refer the interested reader to Appendix \ref{app:Weyl_WKB}.

\section{The Generalized Born Oscillator}
\label{sec:GBO}
In this section, we introduce a one-parameter generalization of the Born oscillator. This modified theory, referred to as the \emph{generalized Born oscillator}, can be viewed as a natural implementation of the Connes cutoff \eqref{eq:C_constraints} on the Berry-Keating Hamiltonian, as will become evident shortly.

The Generalized Born oscillator is determined by the following Hamiltonian
\beq
    H_{\mathrm{GBO}} = \frac{\left(1 + \lambda^u p^{2u}\right)^{\frac{1}{2u}} \left(1 + \lambda^u q^{2u}\right)^{\frac{1}{2u}}}{\lambda}\;,
\label{eq:GBO_Hamiltonian}
\eeq
where we take $\lambda>0$ and $u$ to be a positive integer. This theory possesses a $\mathbb{Z}_4$ symmetry under independent reflections of $p$ and $q$, which allows us to focus on the $p>0$, $q>0$ quadrant.
\begin{figure}[h!]
\centering
\includegraphics[width= 0.6\textwidth]{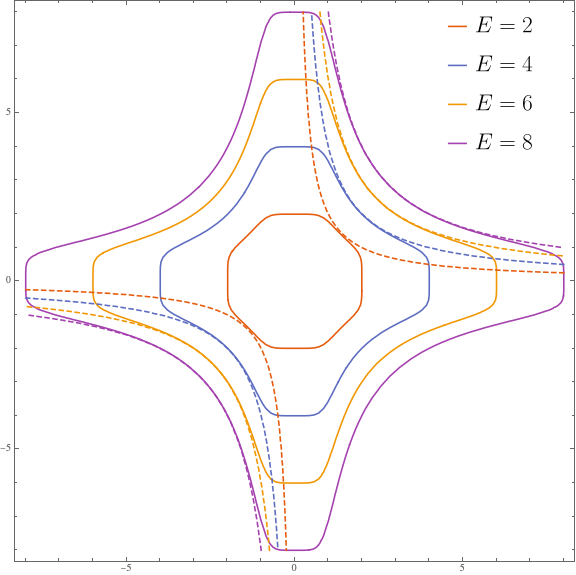}
\caption{\label{fig:GBO_trajectories_energy} Some classical trajectories of the Generalized Born oscillator $H_{\mathrm{GBO}} = E$ for fixed $u=3$ and $\lambda = 1$. The dashed lines are the Berry-Keating trajectories $pq=E$.}
\end{figure}
\begin{figure}[h!]
\centering
\includegraphics[width= 0.6\textwidth]{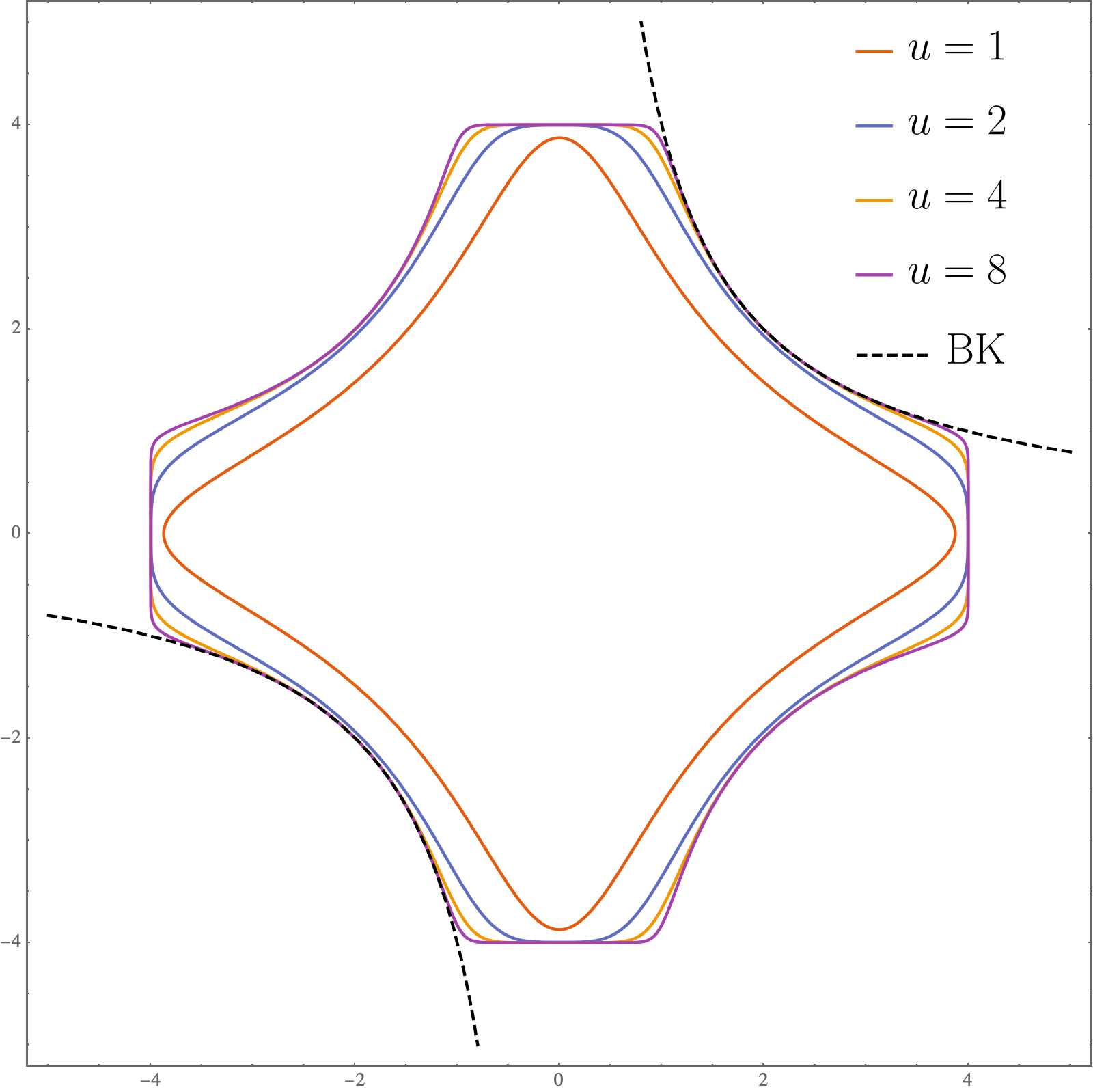}
\caption{\label{fig:GBO_trajectories} Some classical trajectories of the Generalized Born oscillator $H_{\mathrm{GBO}} = E$ for fixed $E=4$ and $\lambda = 1$.}
\end{figure}
The trajectories of this system, depicted in figures \ref{fig:GBO_trajectories_energy} and \ref{fig:GBO_trajectories}, approach the ones of the Berry-Keating Hamiltonian as the energy grows
\beq
    H_{\mathrm{GBO}} \underset{p,q\rightarrow\infty}{\sim} H_{\mathrm{BK}} + \frac{p q}{2u\lambda^u}\left(\frac{1}{p^{2u}} + \frac{1}{q^{2u}}\right) + \cdots\;.
\eeq
Another interesting fact that we can note about the Generalized Born oscillator is that, as $u$ grows towards $\infty$, its trajectories resemble more and more the one obtained with a Connes cutoff from the Berry-Keating ones (see Fig. \ref{fig:pq_lambda}). To be more precise, we can say that in the limit $u\rightarrow\infty$, the trajectories become the following
\beq
    p q = E\;, \qquad  \vert p\vert \leq \sqrt{\lambda} E\;,\quad \vert q \vert \leq \sqrt{\lambda} E\;.
\label{eq:GBO_trajectory_limit}
\eeq
By choosing $\lambda = 1/l_pl_q = 1/(2\pi\hbar)$, the above condition becomes precisely the same as the Connes cutoff prescription, with $\Lambda = \sqrt{\lambda}E$.

Given the above properties, we expect the spectrum of $H_{\mathrm{GBO}}$ to reproduce the Berry-Keating one, with Connes regularization imposed, in the limits $E\rightarrow \infty$ and $u\rightarrow\infty$. Let us look at the first semi-classical order
\beq
    N_{\mathrm{GBO}} + \frac{1}{2} = \frac{1}{2\pi\hbar} \iint\limits_{H_{\mathrm{GBO}}\leq E} dp dq\,+\mathcal{O}(\hbar)\;.
\eeq
As usual, we can evaluate this solving for $p$
\beq
    \begin{split}
        N_{\mathrm{GBO}} + \frac{1}{2} &= \frac{2}{\pi\hbar\lambda} \intop_{0}^{\tilde{q}_t}dq\,\left( \frac{1+\tilde{q}_t^{2u}}{1+q^{2u}} - 1 \right)^{\frac{1}{2u}}  \\
        &= 2\frac{\Gamma\left(1 + \frac{1}{2u}\right)^2}{\Gamma\left(1+\frac{1}{u}\right)} \frac{\tilde{q}_t^2}{\pi\hbar\lambda}\,\phantom{}_2F_1\left(\frac{1}{2u},\frac{1}{2u}, 1 + \frac{1}{u}; -\tilde{q}_t^{2u} \right)\;,
    \end{split}
\eeq
where $\tilde{q}_t = (\lambda^{2u}E^{2u} - 1)^{1/2u}$. Taking the large $E$ limit first, we find
\beq
    N_{\mathrm{GBO}} + \frac{1}{2} = \frac{2E}{\pi\hbar}\left(\log\lambda E - 1\right) - \frac{2E}{\pi \hbar u}\left(\gamma_{\mathrm{E}} + \psi\left(\frac{1}{2u}\right)\right) + \mathcal{O}\left(E^{-2u}\right) + \mathcal{O}\left(\hbar\right)\;,
\eeq
where $\psi(x) = \Gamma'(x)/\Gamma(x)$ is the digamma function. Using the expansion (see \cite{NIST:DLMF}, \S 5.7) $\psi(x)\sim -\gamma_{\mathrm{E}} - 1/x + \mathcal{O}(x)$ we arrive at
\beq
    N_{\mathrm{GBO}} + \frac{1}{2} = \frac{2E}{\pi\hbar}\left(\log\lambda E - 1\right) + \frac{4E}{\pi \hbar} + \mathcal{O}\left(u^{-1}\right) + \mathcal{O}\left(E^{-2u}\right) + \mathcal{O}\left(\hbar\right)\;.
\label{eq:GBO_first_order_quantization}
\eeq
The identifications $N_{\mathrm{GBO}} = 4N_{\mathrm{C}}$ and $\lambda = 1/(2\pi\hbar)$ yield precisely \eqref{eq:C_spectrum_no_lambda} (with the addition of the Maslov index $-1/8$), as expected from the limiting behavior of the trajectories \eqref{eq:GBO_trajectory_limit}. Hence, the Generalized Born oscillator reproduces the first order in the semi-classical expansion of the Berry-Keating spectrum, with the Connes regularization prescription implemented. The difference is that the Hamiltonian \eqref{eq:GBO_Hamiltonian}, having naturally bounded trajectories, allows pushing the semi-classical expansion of the spectrum beyond first approximation to an in principle, arbitrary order in $\hbar$.
\subsection{Quantization of the Generalized Born oscillator}
\label{subsec:QuantizationGBO}
We are going to perform the same analysis we carried on in \S \ref{subsec:QuantizationBO} for the Born oscillator. The quantization condition is expanded in $\hbar$ as in \eqref{eq:quant_cond_for_BO}, where now the coefficients $\Sigma_0(E)$ and $\Sigma_m(E)$ take the following form
\beq
    \Sigma_0(E) = 2\frac{\Gamma\left(1 + \frac{1}{2u}\right)^2}{\Gamma\left(1+\frac{1}{u}\right)} \frac{\tilde{q}_t^2}{\pi\hbar\lambda}\,\phantom{}_2F_1\left(\frac{1}{2u},\frac{1}{2u}, 1 + \frac{1}{u}; -\tilde{q}_t^{2u} \right)\;,
\eeq
and
\beq
    \begin{split}
        \Sigma_m(E) = \frac{C_m\lambda^{2m-1}\tqt^{2-4m}}{(1+\tqt^{2u})^{2m-1}}& \left(P_m(\tqt^{2u}) \, \phantom{}_2F_1\left(1-\frac{2m-1}{2u},1-\frac{2m-1}{2u}, 2 - \frac{2m-1}{u}; -\tilde{q}_t^{2u} \right) \right.\\
        &\left. + Q_m(\tqt^{2u}) \, \phantom{}_2F_1\left(-\frac{2m-1}{2u},1-\frac{2m-1}{2u}, 2 - \frac{2m-1}{u}; -\tilde{q}_t^{2u} \right) \right)\;.
    \end{split}
\label{eq:GBO_sigma_m}
\eeq
Here $P_m(x)$ and $Q_m(x)$ are polynomials of order $2m-1$, the constants $C_m$ have the form
\beq
    C_m = \frac{2^{\frac{2m-1}{u}}}{\sqrt{\pi}} \frac{(2m-1)^{2-4m}}{4u - 2m + 1} \frac{u^{4m-1} \Gamma\left(1-\frac{2m-1}{2u}\right)}{\Gamma\left(\frac{3}{2} - \frac{2m-1}{2u}\right)}\;,
\eeq
and
\beq
    \tqt = \left(\lambda^{2u} E^{2u} - 1\right)^{\frac{1}{2u}}\;.
\eeq
The explicit expression of the polynomials $P_m$ and $Q_m$ for generic values of $u$ rapidly becomes unmanageable as $m$ grows. The simplest case $m=1$ is still rather simple
\beq
    \begin{split}
        P_1(x) &= (4u-1)(2u-1)\frac{(4 u^2 - 2u - 1)x + 2u-1}{48 u^4}\;, \\
        Q_1(x) &= (4u-1)(2u-1)\frac{x - (2u-1)^2}{48 u^4}\;,
    \end{split}
\eeq
and we report the case $m=2$ in Appendix \ref{app:poly}.

In conclusion, the analytical expressions obtained are highly intricate, and as of now, we have not been able to identify a discernible pattern. However, if one is interested in the large $E$ limit, things simplify considerably, as we will see in the next section.
\subsection{The asymptotic behavior of the Generalized  Born oscillator counting function}
It turns out that for the purpose of determining the asymptotic behavior of the counting function, the explicit expressions of the polynomials $P_m$ and $Q_m$ are largely irrelevant.

In fact, in the large $E$ - equivalently, large $\tqt$ - limit of $\Sigma_m$, we observe that
\beq
	\tqt\underset{E\rightarrow\infty}{\sim} \lambda E + \mathcal{O}\left(E^{1-2u}\right)\;,
\eeq
while, from \eqref{eq:GBO_sigma_m} we find
\beq
	\begin{split}
		\Sigma_m(E) \underset{E\rightarrow\infty}{\sim} C_m \lambda^{2m-1}\tqt^{1-2m-2u}&\left[p_{m,2m-1}\left(A\log\tqt + B + \mathcal{O}\left(\tqt^{-2u}\right)\right) \right. \\
		&\left. + q_{m,2m-1}\left(C \tqt^{2u} + D\log\tqt + E + \mathcal{O}\left(\tqt^{-2u}\right)\right) \right]\;,
	\end{split}
\eeq
where $p_{m,2m-1}$ and $q_{m,2m-1}$ are the coefficients of the power $x^{2m-1}$ in the polynomials $P_m(x)$ and $Q_m(x)$, while $A,B,C,D,E$ are constants arising from the large $\tqt$ expansion of the hypergeometric functions in \eqref{eq:GBO_sigma_m}. Thus, we see that for large $E$ and $u\rightarrow\infty$, the expression of $\Sigma_m$ simplifies to
\beq
	\Sigma_m(E) \underset{E,u\rightarrow \infty}{\sim} q_{m,2m-1} C_m C E^{1-2m} + \mathcal{O}\left(E^{1-2m-2u}\log E\right)\;.
\eeq
The constant $C$ is easily found
\beq
	C = \frac{\Gamma\left(2 + \frac{1-2m}{u}\right)}{\Gamma\left(2 + \frac{1-2m}{2u}\right) \Gamma\left(1 + \frac{1-2m}{2u}\right)}\;,
\eeq
and, fortunately, the coefficients $q_{m,2m-1}$ all take a remarkably simple expression\footnote{We verified this up to $m=8$.}
\beq
	q_{m,2m-1} = \mathfrak{q}_m\frac{(4u - 2m + 1)(2u - 2m + 1)}{u^{4m}}\;,
\eeq
where $\mathfrak{q}_m$ are rational numbers. Combining everything, we find that in the limit $u\rightarrow\infty$, the leading term is
\beq
	\Sigma_m(E) \underset{E,u\rightarrow\infty}{\sim} \frac{4}{\pi} \frac{\mathfrak{q}_m}{(2m-1)^{4m-2}} \frac{2^{\frac{1-2m}{u}}}{E^{2m-1}} + \mathcal{O}\left(\frac{\log E}{E^{2u+2m-1}}\right)\;.
\label{eq:GBO_Sigma_limit_form}
\eeq
The correctness of this asymptotic behavior can be verified for larger values of $m$ by fixing $u$ to be a large integer and investigating the leading $E\rightarrow\infty$ behavior of $\Sigma_m$. This additionally allows the determination of the rational numbers $\mathfrak{q}_m$:
\beq
	\begin{split}
		&\mathfrak{q}_1 = \frac{1}{48}\;,\qquad \mathfrak{q}_2 = \frac{567}{640}\;,\qquad \mathfrak{q}_3 = \frac{60546875}{16128}\;,\qquad \mathfrak{q}_4 = \frac{12304904321689}{61440}\;, \\
		&\mathfrak{q}_5 = \frac{2840679949509872253}{45056}\;,\qquad \mathfrak{q}_6 = \frac{10467483340404449525480631647}{134184960}\;\ldots
	\end{split}
\eeq
The final result is that, in the $u\rightarrow\infty$ limit, the large $E$ expansion of quantization condition \eqref{eq:quant_cond_for_BO} for the Generalized Born oscillator is
\beq
    \begin{split}
        n+\frac{1}{2} = &\frac{4E}{\pi\hbar} + \frac{2E}{\pi\hbar}(\log\lambda E - 1) + \frac{\hbar}{12\pi E} + \frac{7\hbar^3}{1440\pi E^3} + \frac{31\hbar^5}{20160\pi E^5}  \\ +&\frac{127 \hbar^7}{430080 \pi E^7} + \frac{511 \hbar^9}{1216512 \pi E^9} + \mathcal{O}\left(E^{-11}\right)+ \mathcal{O}\left(u^{-1}\right)+ \mathcal{O}\left(E^{-2u}\right)\;,
    \end{split}
\eeq
which coincides \underline{almost exactly} with the large $T$ expansion \eqref{eq:sieg_asy} of the mean number $\overline{N(T)}$ of non-trivial zeroes of the Riemann $\zeta$-function, provided we perform the identifications
\beq
    E = \hbar T\;,\qquad \lambda = \frac{1}{2\pi\hbar}\;\qquad n = 4\overline{N(T)} - 4\;.
\eeq
The only ``fly in the ointment'' is the term $4E/(\pi\hbar)$ coming from the expansion of $\Sigma_0(E)$. It is there to remind us that what we are computing here really is a natural regularization of the Connes absorption spectrum \eqref{eq:C_spectrum_no_lambda}.

Finally, it's also important to emphasize that Weyl quantization is a perturbative quantization method with respect to $\hbar$, and therefore, for practical computational reasons, our analytical results are limited to a large but finite order. Nevertheless, the emerging pattern at large $E$ is already highly evident, and it would be exceedingly surprising if the connection with the Riemann-Siegel $\theta$ function were to break down at higher perturbative orders. Just as for the Born oscillator, the Weyl quantization of this modified theory produces a spectrum that reproduces the asymptotic behavior of the mean number of Riemann zeroes $\overline{N(T)}$, together with a number of spurious terms. The difference is that, in the present case, we have an additional parameter $u$. As it turns out, sending $u\rightarrow\infty$ completely cancels out the spurious terms, leaving us purely with the terms reproducing the asymptotic behavior of $\overline{N(T)}$.

\section{Conclusions}\label{sec:conclusions}
The main result of this article is the introduction of a specific regularization scheme for the Berry-Keating Hamiltonian and the investigation of its quantum corrections at very high perturbative orders in $\hbar$, using Weyl quantization. This allowed us to observe the emergence of contributions related to the smooth part of the counting function for the Riemann zeros. In our view, this represents a non-trivial advancement in the exploration of the Berry-Keating proposal, which very ambitiously aims to establish a connection between the simple Hamiltonian $H_{\mathrm{BK}}$ and the Hilbert-Pólya conjecture. As expected from the physical and mathematical considerations briefly outlined in the introductory section, as with previous investigations \cite{Connes1998TraceFI, Berry1999TheRZ, Sierra:2008se, Sierra2011, Sierra_2019, Berry2011}, no direct evidence of a straightforward relationship with prime numbers or the Riemann $\zeta$ function has emerged.

However, let us indulge in some speculative thinking. Firstly, we should acknowledge that the generalized Born oscillator Hamiltonian is defined on a complex multi-sheet Riemann surface, and a pressing question arises: could these characteristics potentially give rise to chaotic behavior in a specific large-$u$ scaling limit? To elaborate a bit further, though still remaining in the realm of speculation, when we delve into the spectral theory of Sturm-Liouville-type problems, we typically encounter two categories of spectral problems. On one hand, there are the ``lateral problems", which involve imposing boundary conditions in two distinct sectors of the complex-$x$ plane (commonly referred to as Stokes sectors) as $|x|$ approaches infinity. Alternatively, one can impose boundary conditions at both the origin and infinity (in some specific sector), resulting in what is known as a ``central spectral problem".

The analysis undertaken in this work can be likened to a lateral problem. On the other hand, findings from a reference like \cite{Berry1999TheRZ} (specifically \S6, page 262, penultimate paragraph) seem to suggest that the Riemann $\zeta$ function might manifest itself as a spectral determinant in what appears to be a spectral central problem (for a comparison, see, for example, \cite{ABCD} or the Appendix B.2 of \cite{Dorey:2006an}), possibly entailing boundary conditions on the multivalued wave functions that involve analytic continuation onto the other Riemann sheets.

In principle, one can envision the emergence of non-trivial spectral determinants in an appropriate scaling limit as the parameter $u$ approaches infinity. Nevertheless, without some additional guiding principles or ideas, the Hamiltonian $H_{\mathrm{GBO}}$ remains one of the infinitely possible multivalued regularizations of $H_{\mathrm{BK}}$, and translating these concepts into concrete progress remains extremely challenging.

Finally, based on our results, an important avenue to explore is establishing a connection with integrable quantum field theories, exact S-matrices, and Bethe Ansatz equations, following the spirit of the ODE/IM correspondence \cite{Dorey:1998pt}. It would be particularly intriguing to uncover the integrable system associated with the generalized Born quantum mechanical model, where local conserved charges on the integrable model's side are related to the high-order WKB-Weyl analysis obtained in this article.

\section*{Acknowledgments}
We would like to express our gratitude to Gianni Coppa for fruitful discussions, during the initial stages of this work and in the period associated with his Master thesis project \cite{CoppaThesis}. Additionally, we are grateful to him for sharing a draft of his article \cite{CoppaArticle}. We also thank Michael Berry, Zoltan Bajnok and Alexander Zamolodchikov for  useful discussions.
This project was partially supported by the INFN project SFT,  grant PHY-2210349, by the Simons Collaboration on Confinement and QCD Strings, and by the FCT Project PTDC/MAT-PUR/30234/2017 ``Irregular connections on algebraic curves and Quantum Field Theory".
S.N. wishes to thank the Department of Physics of the Universit\`{a} degli Studi di Torino for its kind hospitality.
\appendix

\section{Counting functions,  Riemann $\zeta$ and  Riemann-Siegel $\theta$ functions }
\label{sec:RZF}

In order to gain an understanding of the distribution of zeroes and poles of a meromorphic function $f(z)$ inside a region of the complex plane one can employ the argument principle. Let $\gamma$ be a closed path (counterclockwise oriented) bounding the region of interest and $f(z)$ a meromorphic function with $Z$ zeros and $P$ poles\footnote{Counted with their multiplicity and order.}  in such region. Then
\beq
    Z-P=\frac{1}{2\pi i}\oint_\gamma dz \, \frac{f'(z)}{f(z)} \,.
\eeq
As a noteworthy example, if we choose $f(z) = \zeta(z)$ and the region of the complex plane as the \emph{critical strip}:
\beq
    \mathcal{S}_{\epsilon,T} = \left\lbrace s \in \mathbb C\;\Big\vert\; -\e \le \Re s \le 1+\e \;,\quad 0 \le \Im s \le T \right\rbrace\;,
\eeq
remembering that the Riemann zeta has no poles inside this region, we arrive at the following integral expression for the zero \emph{counting function}
\beq
N(T)=\frac{1}{2\pi i}\oint_\gamma ds \, \frac{\zeta'(s)}{\zeta(s)} \,,\qquad \gamma = \partial\mathcal{S}_{\epsilon,T}\;.
\eeq
Thanks to the famous  Riemann's reflection formula, $N(T)$ can be manipulated into  the following form
\beq
    N(T)=\frac{\theta(T)}{\pi} + 1 + \frac{1}{\pi} \Im \log \zeta\(\frac{1}{2}+i T\)
\label{eq:NT}
\eeq
where $\theta(T)$ is the \emph{Riemann–Siegel theta function}
\beq
    \theta (T)=\arg \left[\Gamma \left(\frac{1}{4} + \frac{iT}{2} \right) \right]-{\frac {\log \pi }{2}}T \\
    =\frac {1}{2i}  \log \frac{ \Gamma \(\frac{1}{4}+\frac {iT}{2}\)}{ \Gamma \(\frac{1}{4}-\frac {iT}{2}\) }-\frac {\log \pi }{2}T \,.
\label{eq:sieg}
\eeq

Being a counting function, $N(T)$ is obviously piece-wise constant, while 
\beq
    \overline{N(T)} \doteq \frac{\theta(T)}{\pi} + 1
\label{eq:average_numb_zeroes}
\eeq
can be interpreted as the average number of zeroes. 
The remaining term $N_{\mathrm{fl}}(T) = \pi^{-1}\Im \log \zeta(1/2+i T)$ corresponds then to the fluctuations of $N(T)$ around its mean value $\overline{N(T)}$. Using the well known asymptotic expansion for the $\Gamma$ function, from \eqref{eq:sieg} we readily obtain the large $T$ behavior of the average
\beq
    \begin{split}
        \overline{N(T)} \;\underset{T\rightarrow\infty}{\sim} \;&\frac{T}{2\pi} \left(\log\left(\frac{T}{2\pi}\right) - 1\right) + \frac{7}{8} + \frac{1}{48\pi T} + \frac{7}{5760 \pi T^3} + \frac{31}{80640 \pi T^5} \\
        + & \frac{127}{430080\pi T^7} + \frac{511}{1216512 \pi T^9} + \mathcal{O}\left(T^{-11}\right)\;.
    \end{split}
\label{eq:sieg_asy}
\eeq
This expansion is one of the central expressions for this work. As we will see in later sections, the number of states of certain quantum-mechanical systems exhibit the same behavior \eqref{eq:sieg_asy}. In particular, in \S\ref{sec:GBO} we present a system that reproduces \eqref{eq:sieg_asy}, up to a term linear in $T$.

\section{The Weyl Quantization}\label{app:Weyl_quant}
Here we are going to review the exact quantization approach proposed in \cite{PhysicsPhysiqueFizika.2.131}. This method is inscribed in the framework of the phase-space formulation of quantum mechanics \cite{PhaseSpaceQM} and relies essentially on the \emph{Weyl correspondence} \cite{weyl1927quantenmechanik} between quantum mechanical operators and classical dynamical functions. This approach was proven in \cite{PhysicsPhysiqueFizika.2.131} to provide an exact quantization rule for any quantum-mechanical system with a single degree of freedom and arbitrary Hamiltonian, provided its energy spectrum is non-degenerate. Further, it was shown that at the lowest order in $\hbar$ the exact quantization correctly reproduces the usual Bohr-Sommerfeld rule. We will refer to the approach described here as \emph{Weyl quantization}.

Let us consider a classical Hamiltonian $H(p,q)$ determining the dynamics of a single degree of freedom. Further, let us suppose that we know a rule to consistently associate a quantum-mechanical operator $\hat{H}$ to $H(p,q)$. We will present such a rule momentarily. Then, if the spectrum of $\hat{H}$ is non-degenerate, which we will suppose to be true, it is possible to enumerate the energy eigenvalues $E_n$ using integer numbers $n\in\mathbb{Z}_{\geq 0}$. Now, let $\Theta(x)$ be the Heaviside function\footnote{The value $\Theta(0) = 1/2$ is related to the Maslov index mentioned in \S \ref{sec:BK}, and it is chosen so that the Weyl quantization correctly reproduces the Bohr-Sommerfeld condition at first order in $\hbar$. In principle, different prescriptions could be employed.}
\beq
    \Theta(x) = \left\lbrace \begin{array}{l r}
        \displaystyle 1\;, & x>0 \\[.3cm]
        \displaystyle \frac{1}{2}\;, & x = 0 \\[.5cm]
        \displaystyle 0\;, & x<0
    \end{array}\right.\;.
\eeq
Given a fixed value of $E>0$, the number of states $N(E)$ with energies less than $E$ is simply
\beq
    N(E) = \mathrm{Tr}\,\Theta(E - \hat{H}) = \sum_{m=0}^\infty \Theta(E - E_m)\;.
\label{eq:numb_of_states}
\eeq
For $E = E_n$ this turns into a quantization condition for the energy eigenvalues
\beq
    \mathrm{Tr}\,\Theta(E_n - \hat{H}) = n + \frac{1}{2}\;.
\label{eq:quant_cond}
\eeq

The main idea underlying the Weyl quantization is to express the trace on the left-hand side of \eqref{eq:quant_cond} as an integral over the classical phase space of the system. In order to do so, one can follow the procedure used by E. Wigner in \cite{wigner1997quantum} and define the classical function
\beq
    S(p,q\vert E) = \intop_{-\infty}^{\infty} dx\,e^{i p \frac{x}{\hbar}}\,\left\langle q - x/2 \left\vert \Theta(E - \hat{H}) \right\vert q + x/2 \right\rangle \;.
\label{eq:Weyl_corresp_for_theta}
\eeq
As it is easily verified, the phase space integral of this function yields precisely the trace appearing in \eqref{eq:numb_of_states}
\beq
    \frac{1}{2\pi\hbar} \intop_{\mathbb{R}^2} dpdq\,S(p,q\vert E) = \mathrm{Tr}\,\Theta(E - \hat{H})\;,
\eeq
and, consequently, we can recast the quantization condition \eqref{eq:quant_cond} as a phase space integral
\beq
    \frac{1}{2\pi\hbar} \intop_{\mathbb{R}^2} dpdq\,S(p,q\vert E_n) = n + \frac{1}{2}\;.
\label{eq:quant_cond_phase_space}
\eeq
\subsection{The Weyl correspondence and the Moyal product}

The definition of the classical function \eqref{eq:Weyl_corresp_for_theta} is an instance of the Weyl correspondence. This is a rule that associates uniquely an operator $\hat{A}(\hat{p},\hat{q})$ to a function $A_c(p,q)$ through the following relations
\beq
    \begin{split}
        \hat{A}(\hat{p},\hat{q}) = \intop_{\mathbb{R}^2} d\sigma d\tau\,a(\sigma,\tau) e^{i\hat{p}\sigma + i \hat{q} \tau}\;, \\
        A_c(p,q) = \intop_{\mathbb{R}^2} d\sigma d\tau\, a(\sigma, \tau) e^{i p \sigma + i q \tau}\;.
    \end{split}
\label{eq:Weyl_correspondence}
\eeq
Given that $e^{i p \sigma + i q \tau}/(2\pi)$ form a complete orthonormal set of functions and $\hbar^{1/2}/(2\pi)^{1/2} e^{i p \sigma + i q \tau}$ a complete orthonormal set of operators \cite{groenewold1946principles}, the function $a(\sigma, \tau)$ can be expressed in two equivalent ways
\beq
    a(\sigma, \tau) = \frac{1}{(2\pi)^2} \intop_{\mathbb{R}^2} dp dq\, A_c(p,q) e^{-i p \sigma - i q \tau} = \frac{\hbar}{2\pi} \mathrm{Tr}\,\left[\hat{A}(\hat{p},\hat{q}) e^{-i \hat{p} \sigma - i \hat{q} \tau}\right]\;.
\label{eq:a_function_def}
\eeq
Let us now rewrite the trace on the rightmost side above by using a version of Baker-Campbell-Hausdorff formula
\beq
    e^{\hat{B} + \hat{C}} = e^{\hat{B}/2} e^{\hat{C}} e^{\hat{B}/2}\quad \mathrm{iff}\quad [\hat{B},[\hat{B},\hat{C}]] = [\hat{C},[\hat{B},\hat{C}]] = 0\;.
\eeq
Thanks to this, we have
\beq
    \begin{split}
        a(\sigma, \tau) &= \frac{\hbar}{2\pi} \intop_{-\infty}^{\infty} dq e^{-i q \tau} \left\langle q \left\vert e^{-i \hat{p}\sigma/2} \hat{A}(\hat{p},\hat{q}) e^{- i \hat{p} \sigma/2} \right\vert q \right\rangle  \\
        &= \frac{\hbar}{2\pi} \intop_{-\infty}^{\infty} dq e^{-i q \tau} \left\langle q + \hbar \sigma/2\left\vert \hat{A}(\hat{p},\hat{q})  \right\vert q - \hbar \sigma/2 \right\rangle\;.
    \end{split}
\eeq
Finally, we can plug this expression back into \eqref{eq:Weyl_correspondence}, obtaining
\beq
    A_c(p,q) = \intop_{-\infty}^{\infty} dx \, e^{i p \frac{x}{\hbar}} \left\langle q - x/2\left\vert \hat{A}(\hat{q},\hat{p}) \right\vert q + x/2 \right\rangle\;.
\label{eq:Weyl_correspondence_generic}
\eeq
A property of the Weyl correspondence that we are going to need in the following is that it maps the ordinary product of two operators to the \emph{Moyal product} of classical functions
\beq
    \hat{C} = \hat{A}\hat{B}\quad \Longrightarrow \quad C_c(p,q) = A_c(p,q) \star B_c(p,q) \;,
\eeq
which is defined as \cite{groenewold1946principles}
\beq
    \begin{split}
        A_c(p,q) \star B_c(p,q) &= A_c(p,q)\,\exp\left[i\frac{\hbar}{2} \left(\overset{\leftarrow}{\partial}_p \overset{\rightarrow}{\partial}_q - \overset{\leftarrow}{\partial}_q \overset{\rightarrow}{\partial}_p \right)\right]\,B_c(p,q)  \\
        &= \sum_{n=0}^\infty \frac{1}{n!}\left(i\frac{\hbar}{2}\right)^n\,\sum_{m=0}^{n} (-1)^m \binom{n}{m}  \frac{\partial^n A_c(p,q)}{\partial^{n-m} p\partial^m q} \frac{\partial^n B_c(p,q)}{\partial^m p\partial^{n-m} q}\;.
    \end{split}
\eeq

\subsection{An integral representation for $S$}
Now that we have a Weyl correspondence \eqref{eq:Weyl_correspondence_generic} for generic operators $\hat{A}$ and functions $A_c$, we can return to the expression \eqref{eq:Weyl_corresp_for_theta}. Let us use the following representation of the Heaviside function
\beq
    \Theta(x) = \lim_{\eta\rightarrow 0^+} \frac{i}{2\pi} \intop_{-\infty}^\infty \frac{dz}{z + i\eta} e^{-i z x}\;, 
\eeq
to write
\beq
    S(p,q\vert E) = \lim_{\eta\rightarrow 0^+} \frac{i}{2\pi} \intop_{-\infty}^{\infty} dx\,\intop_{-\infty}^{\infty} \frac{dz}{z + i\eta} \,\left\langle q - x/2 \left\vert e^{-i z (E - \hat{H})} \right\vert q + x/2 \right\rangle\;.
\eeq
Interchanging the integration order and using \eqref{eq:Weyl_correspondence_generic}, we arrive at the integral representation
\beq
    S(p,q\vert E) = \lim_{\eta\rightarrow0^+} \frac{i}{2\pi} \intop_{-\infty}^{\infty} \frac{dz}{z + i\eta} \mathcal{E}_c(p,q\vert z)\;,
\label{eq:S_integral_rep}
\eeq
where the function $\mathcal{E}_c$ is in Weyl correspondence with the operator $\hat{\mathcal{E}} = \exp(- i z(E - \hat{H}))$, which satisfies the differential equation\footnote{The symmetrization of the operators in the differential equation \eqref{eq:Weyl_diff_eq} is needed to recover the usual quantization rule for standard Hamiltonians $H = p^2/2m + V(q)$.}
\beq
    \frac{1}{i}\frac{\partial}{\partial z} \hat{\mathcal{E}} = \frac{1}{2}\left(
\hat{\mathcal{E}} \hat{\mathcal{H}} + \hat{\mathcal{H}} \hat{\mathcal{E}} \right)\;,\qquad \hat{\mathcal{H}} = \hat{H} - E\;.
\label{eq:Weyl_diff_eq}
\eeq
As a consequence and owing to the Weyl correspondence, the classical function $\mathcal{E}_c$ is a solution of the initial value problem
\beq
    \begin{split}
        &\frac{1}{i}\frac{\partial}{\partial z} \mathcal{E}_c(p,q\vert z) = \frac{\mathcal{H}_c(p,q) \star \mathcal{E}_c(p,q\vert z) + \mathcal{E}_c(p,q\vert z) \star \mathcal{H}_c(p,q\vert z)}{2}\;,\\[.2cm]
        &\mathcal{E}_c(p,q\vert 0) = 1\;,
    \end{split}
\label{eq:Weyl_diff_eq_in_z}
\eeq
with $\mathcal{H}_c(p,q) = H(p,q) - E$.

\section{An iterative procedure}
\label{app:iterative}
In order to solve \eqref{eq:Weyl_diff_eq_in_z} and, consequently, explicitly determine the exact quantization condition \eqref{eq:quant_cond_phase_space} through \eqref{eq:S_integral_rep}, we can expand the function $\mathcal{E}_c$ in powers of $\hbar$. In particular, it is convenient to use the following expansion
\beq
    \mathcal{E}_c(p,q\vert z) = e^{iz\mathcal{H}_c(p,q)} \left[1 + \sum_{m=1}^{\infty} G_m(p,q\vert z) \hbar^{2m} \right]\;.
\eeq
Inserting this expansion into \eqref{eq:Weyl_diff_eq_in_z} and massaging the expression, we arrive at a recursion relation for the functions $G_m$
\beq
    \begin{split}
        &\frac{1}{i}\frac{\partial}{\partial z} G_m(p,q\vert z) = \sum_{n=1}^m \frac{(-2)^{-2n}}{(2n)!} \sum_{k=0}^{2n} (-1)^k \binom{2n}{k} e^{-i z H(p,q)} \\ 
        &\phantom{\frac{1}{i}\frac{\partial}{\partial z} G_m(p,q\vert z)} \times \frac{\partial^{2n}}{\partial p^{2n-k}\partial q^k} \left(e^{i z H(p,q)} G_{m-n}(p,q\vert z)\right) \frac{\partial^{2n}H(p,q)}{\partial p^k\partial q^{2n-k}}\;.
    \end{split}
\label{eq:recursion_relation}
\eeq
This relation allows us to determine $\mathcal{E}_c(p,q\vert z)$ at order $\mathcal{O}(\hbar^n)$ in terms of the same function at order $\mathcal{O}(\hbar^{n-1})$. We can then expand the function $S$ in powers of $\hbar$
\beq
    S(p,q\vert E) = \Theta\Big( E - H(p,q) \Big) + \sum_{m=1}^{\infty} S_m(p,q\vert E) \hbar^{2m}\;,
\eeq
where
\beq
    S_m(p,q\vert E) = \lim_{\eta\rightarrow 0^+} \frac{i}{2\pi} \intop_{-\infty}^{\infty} dz\, \frac{e^{iz\mathcal{H}_c(p,q)}}{z+i\eta} G_m(p,q\vert z)\;.
\eeq
Looking closely at \eqref{eq:recursion_relation} we realize that the functions $G_m$ must be polynomials in $z$, of lowest order $2$ and highest $3m$
\beq
    G_m(p,q\vert z) = \sum_{\ell=2}^{3m} g_{m,\ell}(p,q) z^{\ell}\;.
\label{eq:G_as_poly}
\eeq
Thus, using the identity
\beq
    \lim_{\eta\rightarrow0^+} \frac{i}{2\pi} \intop_{-\infty}^{\infty} dz\,\frac{e^{iz\mathcal{H}_c(p,q)}}{z+\eta}\,z^{\ell} = \left(i\frac{d}{dE}\right)^{\ell} \Theta\Big(E - H(p,q)\Big)\;,
\eeq
we can write
\beq
    S_m(p,q\vert E) = i\sum_{\ell = 2}^{3m} g_{m,\ell}(p,q)\, \left(i\frac{d}{dE}\right)^{\ell-1} \delta\Big(E - H(p,q)\Big)\;,
\eeq
where we used the identity $d/dx\,\Theta(x) = \delta(x)$.

Collecting everything we have derived, we can write the quantization condition \eqref{eq:quant_cond_phase_space} as
\beq
    n+\frac{1}{2} = \frac{1}{\hbar}\Sigma_0(E) + \sum_{m=1}^\infty \Sigma_m(E) \hbar^{2m-1}\;,
\label{eq:quant_exp_in_h}
\eeq
where
\beq
    \begin{split}
        \Sigma_0(E) &= \frac{1}{2\pi}\intop_{\mathbb{R}^2}dpdq\,\Theta\Big(E - H(p,q)\Big)\;, \\
        \Sigma_m(E) &= \frac{i}{2\pi} \sum_{\ell = 2}^{3m} \left(i\frac{d}{dE}\right)^{\ell-1} \intop_{\mathbb{R}^2} dpdq\, g_{m,\ell}(p,q) \delta\Big(E - H(p,q)\Big)\;.
    \end{split}
\eeq
We can further simplify these expressions by changing integration variables from $(p,q)$ to $(H,q)$, inverting the expression $H = H(p,q)$ in favor of $p = p(H,q)$. One need to be careful about multivaluedness of $p$ as a function of $H$. In the case considered in the main text, we can exploit the $\mathbb{Z}_4$ symmetry of the systems to focus on the region $p\geq0$ and $q\geq0$. Then we can see that $p(H,q)$ is non-negative only for $0\leq q \leq q_t(H)$, with $q_t(H)$ being the positive turning point, such that $H(0,q_t(H)) = 0$. With this restriction, we can change variables
\beq
    \begin{split}
        \Sigma_0(E) &= \frac{2}{\pi} \intop_{0}^{q_t(E)} dq\,\intop_{0}^{\infty}dH\,\frac{dp(H,q)}{dH}\,\Theta\Big(E - H\Big) = \frac{2}{\pi} \intop_{0}^{q_t(E)} dq\, \Big[p(E,q) - p(0,q)\Big]\;, \\
        \Sigma_m(E) &= \frac{2i}{\pi}\sum_{\ell = 2}^{3m} \left(i\frac{d}{dE}\right)^{\ell - 1} \intop_0^{q_t(E)} dq\,\intop_0^{\infty} dH\, \frac{dp(H,q)}{dH} g_{m,\ell}(p(H,q),q)\delta\Big(E - H\Big)  \\
        & = \frac{2i}{\pi}\sum_{\ell = 2}^{3m} \left(i\frac{d}{dE}\right)^{\ell - 1} \intop_0^{q_t(E)} dq\, \frac{dp(E,q)}{dE} g_{m,\ell}(p(E,q),q)\;.
    \end{split}
\label{eq:Sigma_eq_final}
\eeq
To recapitulate, we can expand the quantization condition around $\hbar = 0$ and compute the coefficients $\Sigma_m(E)$ using \eqref{eq:Sigma_eq_final}. This requires us to extract the functions $g_{m,\ell}(p,q)$, which are the coefficients of the polynomials $G_m(p,q\vert z)$ \eqref{eq:G_as_poly}. These have to be determined from the recursion equation \eqref{eq:recursion_relation}. All these steps can be automated in a Mathematica\textsuperscript{\textcopyright} script, providing us with an iterative routine to determine the quantization condition \eqref{eq:quant_cond_phase_space} up to, in principle, any desired order $\mathcal{O}\left(\hbar^k\right)$. In practice, for the Hamiltonians considered in the main text, the computations involved in the routine become excessively taxing for large $k$, so we limited ourselves to $k=11$ for the Born oscillator and $k=5$ for the modified one. For the latter, however, it is not necessary to determine the exact form of $\Sigma_{m}(E)$ to determine the order $\hbar^{2m-1}$ contribution to the quantization condition in the large $E$ and $u$ limits. In fact, as explained in the main text \eqref{eq:GBO_Sigma_limit_form}, this limit is controlled by a set of rational numbers $\mathfrak{q}_m$, whose value is independent of $u$. For this reason, they can be computed by fixing $u$ to be some natural number, which greatly improves the speed of the procedure presented here and allowed us to obtain the large $E$ and $u$ behavior of the quantization condition up to order $\mathcal{O}\left(\hbar^{11}\right)$.

\section{Weyl vs WKB quantizations}
\label{app:Weyl_WKB}
In this Appendix, we are going to show that the Weyl quantization amounts to the choice of a very specific ordering prescription for the quantum Hamiltonian. As such, we expect the spectrum extracted via the procedure detailed in Appendix \ref{app:iterative} to coincide with the one obtained from the usual WKB expansion. This fact is easily proven true for Hamiltonians of the form $H = \frac{p^2}{2m} + V(q)$, where no ordering issue arises. We are going to provide evidence that the identity of the spectra holds true also for more complicated classical Hamiltonians of product form $H = F(p)G(q)$, by computing the WKB quantization condition up to order $\mathcal{O}\left(\hbar^5\right)$ for the Born oscillator \eqref{eq:BO_Hamiltonian} and comparing it to the one found with the Weyl quantization procedure (\ref{eq:app_BO_Sigma} -- \ref{eq:app_BO_poly_last}).

\subsection{Operator ordering from Weyl quantization}

From the expressions \eqref{eq:Weyl_correspondence}, it is possible to calculate the action of the operator $\hat{A}$ on a generic function $f(q)$ with sufficient fast decay at $\vert q \vert \rightarrow \infty$
\beq
    \begin{split}
        \hat{A}(\hat{p},\hat{q}) f(q) &= \intop_{\mathbb{R}^2} d\sigma d\tau\,a(\sigma, \tau) e^{-i\hbar\frac{\sigma\tau}{2}} e^{i\tau\hat{q}} e^{i\sigma\hat{p}} f(q)  \\
        & = \intop_{\mathbb{R}^2} d\sigma d\tau\,a(\sigma, \tau) e^{i\hbar\frac{\sigma\tau}{2}} e^{i\tau q} f(q + \hbar\sigma)\;.
    \end{split}
\eeq
Then, using the definition \eqref{eq:a_function_def} of the function $a(\sigma,\tau)$, we find
\beq
    \begin{split}
        \hat{A}(\hat{p},\hat{q}) f(q) & = \frac{1}{(2\pi)^2} \intop_{\mathbb{R}^2} dq'dp'\,\intop_{\mathbb{R}^2} d\sigma d\tau\, A_c(p',q') e^{-i \sigma p'} e^{-i \tau (q' - q - \hbar \sigma/2)} f(q + \hbar \sigma)  \\
        &= \frac{1}{2\pi} \intop_{\mathbb{R}^2} dp d\sigma\, A_c(p,q + \hbar \sigma/2) e^{-i \sigma p} f(q + \hbar \sigma)\;.
    \end{split}
\label{eq:Weyl_operator_action}
\eeq
The formula above defines the quantum operator $\hat{A}$ in position representation. We can then use it to perform a WKB expansion of the associated spectrum.

In order to gain a more concrete feel for the formula \eqref{eq:Weyl_operator_action}, let us make some examples in order of increasing complexity:
\begin{enumerate}
    \item Functions of position: $A_c(p,q) = V(q)$.\\
    This is a trivial case
    \beq
        \hat{A}(\hat{p},\hat{q}) f(q) = \frac{1}{2\pi} \intop_{\mathbb{R}^2} dp d\sigma\, e^{-i \sigma p} f(q + \hbar\sigma) V(q + \hbar \sigma/2) = f(q) V(q)\;,
    \eeq
    which naturally agrees with the canonical quantization prescription.
    \item Monomials in momentum: $A_c(p,q) = p^n$.\\
    Here we use simple manipulations to find
    \beq
        \begin{split}
            \hat{A}(\hat{p},\hat{q}) f(q) &= \frac{1}{2\pi} \intop_{\mathbb{R}^2} dp d\sigma\, f(q + \hbar\sigma) \left(i\frac{d}{d\sigma}\right)^n e^{-i \sigma p}  \\
            &= \frac{1}{2\pi} \intop_{\mathbb{R}^2} dp d\sigma\, e^{-i \sigma p} \left(-i\frac{d}{d\sigma}\right)^n f(q + \hbar\sigma) \\
            & = \left(-i\hbar\frac{d}{dq}\right)^n f(q) = \hat{p}^n f(q)\;,
        \end{split}
    \eeq
    which, again, is consistent with the canonical quantization.
    \item The Berry-Keating Hamiltonian: $A_c(p,q) = p q$.\\
    Things become more interesting in this case. Using the same manipulations as above, we see that
    \beq
        \begin{split}
            \hat{A}(\hat{p},\hat{q}) f(q) &= \frac{1}{2\pi} \intop_{\mathbb{R}^2} dp d\sigma\, (q + \hbar\sigma/2) f(q + \hbar\sigma) \left(i\frac{d}{d\sigma}\right) e^{-i \sigma p}   \\
            & = \frac{1}{2\pi} \intop_{\mathbb{R}^2} dp d\sigma\, e^{-i \sigma p} \left(-i\frac{d}{d\sigma}\right) (q + \hbar\sigma/2) f(q + \hbar\sigma)  \\
            & = -i\hbar\left(qf'(q) + \frac{1}{2} f(q) \right) = \frac{\hat{q}\hat{p} + \hat{p}\hat{q}}{2}f(q)\;.
        \end{split}
    \eeq
    The Weyl correspondence produces the symmetrized version of the naive quantization $pq\,\to\,\hat{p}\hat{q}$.
    \item Operators of factorized form: $A_c(p,q) = F(p)G(q)$.\\
    For this case, we further suppose that the function $F(p)$ is expandable in Taylor series
    \beq
        F(p) = \sum_{n=0}^{\infty} F_n p^n\;.
    \eeq
    Then we can evaluate the integral as done before, finding
    \beq
        \begin{split}
            \hat{A}(\hat{p},\hat{q}) f(q) &= \sum_{n=0}^\infty i^n \frac{F_n}{2\pi} \intop dp d\sigma\, G(q + \hbar \sigma/2) f(q + \hbar \sigma) \frac{d^{n}}{d\sigma^{n}} e^{-i\sigma p}  \\
            &= \sum_{n=0}^\infty (-i)^n F_n \intop d\sigma\, \delta(\sigma) \frac{d^{n}}{d\sigma^{n}} G(q + \hbar \sigma/2) f(q + \hbar \sigma) \\
            &= \sum_{n=0}^\infty (-i\hbar)^n F_n \sum_{m=0}^{n} \frac{1}{2^m} \binom{n}{m} G^{(m)}(q)f^{(n-m)}(q)  \\
            &= \sum_{n=0}^\infty F_n \sum_{m=0}^{n} \frac{1}{2^m} \binom{n}{m} \hat{p}^m G(q) \hat{p}^{n-m} f(q) \;.
        \end{split}
    \label{eq:Weyl_operator_action_prod}
    \eeq
    Again, we see that the Weyl quantization yields a quantum operator with a very specific ordering of the operators $\hat{p}$ and $\hat{q}$.
\end{enumerate}

\subsection{WKB analysis of Born oscillator}

The Born oscillator Hamiltonian \eqref{eq:BO_Hamiltonian} is of factorized form
\beq
    H_{\mathrm{BO}}(p,q) = F(p)F(q)\;,\qquad F(x) = \sqrt{1+\lambda x^2} = \sum_{n=0}^{\infty} \binom{1/2}{n} \lambda^n x^{2n}\;.
\eeq
According to the analysis above, the corresponding operator obtained via the Weyl quantization acts of functions of $q$ as
\beq
    \hat{H}_{\mathrm{BO}}(\hat{p},\hat{q})f(q) = \sum_{n=0}^{\infty}(-1)^n \binom{1/2}{n} \hbar^{2n} \lambda^n \sum_{m=0}^{2n} \frac{1}{2^m}\binom{2n}{m}F^{(m)}(q)\,f^{(2n-m)}(q)\;.
\label{eq:weyl_bo_action}
\eeq
Now, let us make the WKB ansatz
\beq
    f(q) = \exp\left[\frac{i}{\hbar} \sum_{\ell = 0}^{\infty} S_{\ell}(q) \hbar^{\ell} \right]\;,
\eeq
and expand the eigenvalue equation
\beq
    \hat{H}_{\mathrm{BO}}(\hat{p},\hat{q}) f(q) = E f(q)\;,
\label{eq:app_eig_eq}
\eeq
around $\hbar = 0$. We will limit ourselves to order $\hbar^4$. Let us introduce the functions $d_{2n,m}(q)$ from the expansion
\beq
    (-1)^n\hbar^{2n}\frac{f^{(2n)}(q)}{f(q)} = \sum_{m=0}^{\infty} d_{2n,m}(q)\hbar^m\;.
\label{eq:app_d_coeff_def}
\eeq
Then the expansion for each fixed $n$ of \eqref{eq:weyl_bo_action} is
\beq
    \begin{split}
        \binom{1/2}{n}\lambda^n&\left[F  d_{2n,0} + \left(F d_{2n,1} + \frac{n}{i}F' d_{2n-1,0} \right)\hbar   \right.  \\
        &+ \left(F d_{2n,2} + \frac{n}{i}F' d_{2n-1,1} - n\frac{2n-1}{4} F''  d_{2n-2,0} \right)\hbar^2 \\
        &+ \left(F d_{2n,3} + \frac{n}{i}F' d_{2n-1,2} - n\frac{2n-1}{4} F''  d_{2n-2,1} \right.  \\
        &- \left. \frac{n}{i}\frac{(2n-1)(n-1)}{12} F^{(3)}  d_{2n-3,0} \right)\hbar^3  \\
        &+  \left(F d_{2n,4} + \frac{n}{i}F' d_{2n-1,3} - n\frac{2n-1}{4} F''  d_{2n-2,2}  \right. \\
        &- \frac{n}{i}\frac{(2n-1)(n-1)}{12} F^{(3)}  d_{2n-3,1}  \\
        &\left. \left. +n\frac{(2n-1)(n-1)(2n-3)}{96} F^{(4)}  d_{2n-4,0} \right)\hbar^4 \right] + \mathcal{O}\left(\hbar^5\right)\;.
    \end{split}
\eeq
Now we can perform the sum over $n$ and compare the two sides of the eigenvalue equation \eqref{eq:app_eig_eq} order by order in $\hbar$. Using the definition \eqref{eq:app_d_coeff_def} of the functions, $d_{2n,m}(q)$ we can finally derive an expression for $S_l'(q)$ with $l=0,1,2,3,4$. Sparing the technical manipulations, the results are
\beq
    \begin{split}
        S_0'(q) &= \frac{\sqrt{\tqt^2 - \lambda q^2}}{\sqrt{\lambda}\sqrt{1+\lambda q^2}}\;, \\
        S_2'(q) &= \frac{\lambda^{3/2}}{8(1 + \tqt^2)} \frac{(1 + \lambda q^2)^3 - (1 + \tqt^2) (1 + \lambda q^2) (4 + \lambda q^2) + (1 + \tqt^2)^2 (3 + 5 \lambda q^2)}{(\tqt^2 - \lambda q^2)^{5/2}\sqrt{1 + \lambda q^2}}\;, \\
        S_4'(q) &= -\frac{\lambda^{7/2}}{128 (1 + \tqt^2)^3} \frac{1}{(\tqt^2 - \lambda q^2)^{11/2}(1 + \lambda q^2)^{3/2}} \\
        &\times \left[ 4 (1 + \lambda q^2)^6 (11 + \lambda q^2 (58 + 31 \lambda q^2))  \right. \\
        &- (1 + \tqt^2) (1 + \lambda q^2)^5 (263 + \lambda q^2 (1334 + 703 \lambda q^2))  \\
        &+ 2 (1 + \tqt^2)^2 (1 + \lambda q^2)^4 (326 + \lambda q^2 (1595 + 829 \lambda q^2)) \\
        &- (1 + \tqt^2)^3 (1 + \lambda q^2)^3 (826 + \lambda q^2 (4264 + 2021 \lambda q^2)) \\
        & + 2 (1 + \tqt^2)^4 (1 + \lambda q^2)^2 (306 + \lambda q^2 (1273 + 933 \lambda q^2)) \\
        &+ (1 + \tqt^2)^5 (1 + \lambda q^2) (-303 + \lambda q^2 (-666 + 17 \lambda q^2))  \\
        &+ \left. (1 + \tqt^2)^6 (84 + 4 \lambda q^2 (74 + 41 \lambda q^2)) \right]\;.
    \end{split}
\eeq
The functions $S_{2l-1}'(q)$ are total derivatives, as expected from the form of the quantization condition \eqref{eq:quant_exp_in_h}
\beq
    \begin{split}
        S_1'(q) &= i \frac{\lambda q}{2} \frac{\tqt^2 - 1 - 2 \lambda q^2}{(\tqt^2 - \lambda q^2)(1 + \lambda q^2)} = \frac{i}{4}\frac{d}{dq} \log\left[\left(\tqt^2 - \lambda q^2\right) \left(1 + \lambda q^2\right)\right]\;, \\
        S_3'(q) &= i\frac{\lambda^2}{16}\frac{d}{dq} \left( \frac{2\lambda q^2}{(1 + \tqt^2)^2} + \frac{\tqt^4 + 2 \tqt^2 (1 + \lambda q^2) + \lambda q^2 (3 + 2 \lambda q^2)}{(\tqt^2 - \lambda q^2)^3} \right)\;.
    \end{split}
\eeq
Correctly, $S_1'(q)$ produces the $1/2$ in the left-hand side of \eqref{eq:quant_exp_in_h}, while all the functions $S_{2l-1}'(q)$ with $l>1$ yield a vanishing contribution.
All that is left to do is perform the integration around the cut between the two classical turning points $q = \pm \tqt/\sqrt{\lambda}$. Doing so for the above expression we obtain \emph{precisely} the same expressions \eqref{eq:BO_sigma_0} and (\ref{eq:BO_sigma_general_form}, \ref{eq:BO_sigma_elliptic_coefficients_1}, \ref{eq:BO_sigma_elliptic_coefficients_2}).

\section{Expansion of the quantization condition}\label{app:poly}
Here we display the expressions of the functions $\Sigma_0$ and $\Sigma_m$ in \eqref{eq:quant_exp_in_h} for both the Born oscillator \eqref{eq:BO_Hamiltonian} and the Generalized one \eqref{eq:GBO_Hamiltonian}.

\subsection{The Born oscillator}
\label{app:poly_BO}
The expression for $\Sigma_0(E)$ is easily obtained by evaluating the area contained inside the curve $H_{\mathrm{BO}}(p,q) = E$ in phase space. The result is
\beq
    \Sigma_0(E) = 2\frac{\left(1 + \tqt^2\right)\mathsf{K}\left( -\tqt^2\right) - \mathsf{E}\left( -\tqt^2\right)}{\pi\lambda}\;,\qquad \tqt = \sqrt{\lambda^2 E^2 - 1}\;.
\eeq
For $m>0$, we find the following general form
\beq
    \Sigma_m(E) = \lambda^{2m-1}\frac{P_m(\tqt)\mathsf{K}\left(-\tqt^2\right) + Q_m(\tqt)\mathsf{E}\left(-\tqt^2\right)}{\pi \tqt^{4m-2}\left(1+ \tqt^2\right)^{2m-1}}\;,
\label{eq:app_BO_Sigma}
\eeq
with $P_m(x)$ and $Q_m(x)$ polynomials of order $3m-2$. Their expressions for $m\leq 5$ are as follows
\beq
    \begin{split}
        P_1(x) &= \frac{1 + x}{12} \\
        Q_1(x) &= \frac{-1 + x}{12}\;,
    \end{split}
\eeq
\beq
    \begin{split}
        P_2(x) &= -\frac{56 + 233x + 363x^2 + 5323 x^3 + 2257x^4}{2880} \\
        Q_2(x) &= \frac{56 + 105x + 164x^2 + 5233 x^3 + 14x^4}{2880}\;,
    \end{split}
\eeq
\beq
    \begin{split}
        P_3(x) = \frac{1}{161280}&\Big(3968 + 24488x + 63417x^2 + 88604x^3 \\
        &+ 69614x^4 - 4995660x^{5} - 1641631x^{6} + 152336x^{7} \Big) \\
        Q_3(x) = \frac{-1}{161280}&\Big(3968 + 22504x + 52413x^2 + 63680x^3 \\
        &+ 40426x^4 - 4931256x^{5} + 411089x^{6} - 248x^{7} \Big)\;,
    \end{split}
\eeq
\beq
    \begin{split}
        P_4(x) = \frac{-1}{5160960}&\Big(390144 + 3249280x + 11949984x^2 + 25450111x^3 \\
        &+ 34524870x^4 + 30655989x^{5} + 18659436x^{6} \\
        &+ 7336600121x^{7} + 3680016390x^{8} + 271444211x^{9}  \\
        &+ 30495048x^{10}\Big)\\
        Q_4(x) = \frac{1}{5160960}&\Big(390144 + 3054208x + 10447264x^2 + 20405175x^3 \\
        &+ 24887603x^4 + 19216558x^{5} + 10138454x^{6} \\
        &+ 7264577571x^{7} + 420064951x^{8} + 109726120x^{9}  \\
        &+ 6096x^{10}\Big) \;,
    \end{split}
\eeq
\beq
    \begin{split}
        P_5(x) = &\frac{1}{1362493440}\Big(586055680 + 6170450944x + 29505390720x^2 \\
        &+ 84615632480x^3 + 161803832495x^4 + 216819152637x^{5} \\
        &+ 208320760071x^{6} + 145960641489x^{7} + 67944358185x^{8}  \\
        &- 186959663282761x^{9} - 142784617345267x^{10} \\
        &- 27482434315173x^{11} - 2208700224428x^{12} \\
        &- 23790892160x^{13}\Big)\\
        Q_5(x) = &\frac{-1}{1362493440}\Big(586055680 + 5877423104x + 26603307648x^2 \\
        &+ 71663003360x^3 + 127463100631x^4 + 156844408203x^{5} \\
        &+ 136079957703x^{6} + 84851637291x^{7} + 30888603237x^{8}  \\
        &- 186400257077383x^{9} - 53425800506507x^{10}  \\
        &- 6870155121039x^{11} - 83250817016x^{12} - 2289280x^{13}\Big) \;.
    \end{split}
\label{eq:app_BO_poly_last}
\eeq

\subsection{The Generalized Born oscillator}
We will use the same notation as in \ref{app:poly_BO}. As usual the first term is obtained by computing the phase space area contained inside the classical trajectories
\beq
    \Sigma_0(E) = 2\frac{\Gamma\left(1 + \frac{1}{2u}\right)^2}{\Gamma\left(1+\frac{1}{u}\right)} \frac{\tilde{q}_t^2}{\pi\hbar\lambda}\,\phantom{}_2F_1\left(\frac{1}{2u},\frac{1}{2u}, 1 + \frac{1}{u}; -\tilde{q}_t^{2u} \right)\;.
\eeq
The higher coefficients $\Sigma_m(E)$ have the general form
\beq
    \begin{split}
        \Sigma_m(E) = \frac{C_m\lambda^{2m-1}\tqt^{2-4m}}{(1+\tqt^{2u})^{2m-1}}& \left(P_m(\tqt^{2u}) \, \phantom{}_2F_1\left(1-\frac{2m-1}{2u},1-\frac{2m-1}{2u}, 2 - \frac{2m-1}{u}; -\tilde{q}_t^{2u} \right)  \right.\\
        &\left. + Q_m(\tqt^{2u}) \, \phantom{}_2F_1\left(-\frac{2m-1}{2u},1-\frac{2m-1}{2u}, 2 - \frac{2m-1}{u}; -\tilde{q}_t^{2u} \right) \right)\;,
    \end{split}
\eeq
where $P_m(x)$ and $Q_m(x)$ are polynomials of order $2m-1$ and
\beq
    C_m = \frac{2^{\frac{2m-1}{u}}}{\sqrt{\pi}} \frac{(2m-1)^{2-4m}}{4u - 2m + 1} \frac{u^{4m-1} \Gamma\left(1-\frac{2m-1}{2u}\right)}{\Gamma\left(\frac{3}{2} - \frac{2m-1}{2u}\right)}\;.
\eeq
Here, for completeness, we report the explicit polynomials for $m=1,2$. We limit ourselves to these cases, since for higher values of $m$ their expression is practically unmanageable
\beq
    \begin{split}
        P_1(x) &= (4u-1)(2u-1)\frac{(4 u^2 - 2u - 1)x + 2u-1}{48 u^4}\;, \\
        Q_1(x) &= (4u-1)(2u-1)\frac{x - (2u-1)^2}{48 u^4}\;,
    \end{split}
\eeq
\beq
    \begin{split}
        P_2(x) &= -81 (4 u-3) \frac{2 u \left(2 u \left(8 u (3 u+2) \left(4 u \left(u^2+u+1\right)+7\right)-125\right)-205\right)+99}{1280 u^8} x^3  \\
        &- 81 \frac{16 u (u (4 u (u (u (u (4 u (76 u - 41) + 571) - 788) - 227) + 358) - 405) + 180) - 513}{1280 u^8} x^2 \\
        &- 81 \frac{2 u (2 u (4 u (u (2 u (2 u (32 u (2u + 33) - 1155) + 69) - 879) + 927) - 163) + 33) - 135}{1280 u^8} x  \\
        &- 243 \frac{(4 u-3) (8 u-1) (4 (u-2) u+3)^2}{1280 u^8}\;, \\
        Q_2(x) &= 567 \frac{(4 u-3) (2 u-3)}{640 u^8} x^3 \\
        &+ 81 \frac{2 u (4 u (8 u (u (u (u (2 u (64 u - 17) + 15) + 54) - 242) - 44) + 1185) - 1335) + 315}{1280 u^8} x^2 \\
        &+ 81\frac{2 u (u (2 u (2 u (4 u (2 u (u (32u + 153) + 10) - 777) + 1763) - 663) + 589) - 147) + 27}{320 u^8} x \\
        &+ 81 \frac{(1-2u)^2 (4 u-3) (8 u-1) (2 u-3)^3}{1280 u^8}\;.
    \end{split}
\eeq

\newpage

\bibliography{biblio3}

\end{document}